\documentclass[11pt,oneside,a4paper]{article}

\usepackage[margin=2.5cm]{geometry}

\usepackage{setspace}
\usepackage{soul}
    \setul{0.5ex}{0.3ex}

\usepackage{amsfonts,dsfont}
\usepackage{amsmath,amssymb}
\usepackage{mathrsfs}

\usepackage{amsthm}
\theoremstyle{definition}
\newtheorem{definition}{Definition}[section]
\newtheorem{lemma}{Lemma}[section]

\usepackage[square,numbers]{natbib}

\usepackage{mathtools}
\usepackage{multirow}

\usepackage[titletoc,title]{appendix}

\usepackage{lipsum}
\usepackage[ruled]{algorithm2e}

\SetAlFnt{\small}
\SetAlCapFnt{\small}
\SetAlCapNameFnt{\small}
\SetAlCapHSkip{0pt}
\IncMargin{-\parindent}

\usepackage{graphicx}
\graphicspath{ {./} }
\DeclareGraphicsExtensions{.pdf}

\title{\textbf{Optimizing the Convergence Rate of the Quantum Consensus: A Discrete Time Model}}
\author{Saber Jafarizadeh  \\ {saber.jafarizadeh@sydney.edu.au} }
\date{}
\providecommand{\keywords}[1]{\textbf{\textit{Index terms---}} #1}

\usepackage{fancyhdr}

\begin{document}

\maketitle

\markboth{S. Jafarizadeh}    {Discrete Time Quantum Consensus}

\bibliographystyle{ieeetran}

\begin{abstract}

Motivated by the recent advances in the field of quantum computing, quantum systems are modelled and analyzed as networks of decentralized quantum nodes which employ distributed quantum consensus algorithms for coordination.
In the literature, both continuous and discrete time models have been proposed for analyzing these algorithms.
This paper aims at optimizing the convergence rate of the discrete time quantum consensus algorithm over a quantum network with $N$ qudits.
The induced graphs are categorized in terms of the partitions of integer $N$ by arranging them as the Schreier graphs.
It is shown that the original optimization problem reduces to optimizing the Second Largest Eigenvalue Modulus (SLEM) of the weight matrix.
Exploiting the Specht module representation of partitions of $N$, the Aldous' conjecture is generalized to all partitions (except ($N$)) in the Hasse diagram of integer $N$.
Based on this result, it is shown that the spectral gap of Laplacian of all induced graphs corresponding to partitions (other than ($N$)) of $N$ are the same, while the spectral radius of the Laplacian is obtained from the feasible least dominant partition in the Hasse diagram of integer $N$.
The semidefinite programming formulation of the problem is addressed analytically for $N \leq d^2 + 1$ and a wide range of topologies where closed-form expressions for the optimal results are provided.
For a quantum network with complete graph topology, solution of the optimization problem based on group association schemes is provided for all values of $N$.

\end{abstract}


\section{Introduction}

Consensus is essential to the coordinated control of dynamical systems modeled as networks of autonomous agents, such as power grids and social networks \cite{WeiBookVehicleConsensus2008,KocarevBookComplexConsensus2013}.
Reaching consensus as a cooperative collective behaviors in networks of autonomous agents have been studied extensively in the context of distributed control and optimization on networks \cite{Olfati2004,Jadbabaie2003,Tsitsiklis1986}.
Optimization of the discrete time model of the classical distributed average consensus algorithm has been addressed analytically in \cite{SaberThesis2015,SaberSensorJournal2011,SaberICC2011}.

In the recent advances in the field of quantum distributed computing \cite{Broadbent2008,Buhrman2003,Denchev:2008}, quantum systems are analyzed as networks of Quantum nodes that coordinate and carry out computation without any centralized observation.
In the literature, both continuous time and discrete time models have been considered for analysis of the consensus algorithm over quantum networks.
In \cite{Petersen2015IEEETranAutControl,Petersen2015ACCPartI,Petersen2015ACCPartII,SaberContQuanArXiv}, authors consider the continuous time model of the classical consensus dynamics where their approach is based on the induced graphs of the quantum interaction graph.
They establish necessary and sufficient conditions for exponential and asymptotic quantum consensus, respectively, for switching quantum interaction graphs.
In \cite{SaberContQuanArXiv}, the convergence rate of the continuous time quantum consensus is optimized and it is shown that the optimal convergence rate is independent of the value of $d$ in qudits.
Authors in \cite{PetersenRef15,MazzarellaCDC2013,MazzarellaArXiv} employ the discrete time model of the classical consensus algorithm to addressed the consensus in quantum networks.
They reinterpret the quantum consensus algorithm as a symmetrization problem, and they derive the general conditions for convergence.

In this paper we optimize the convergence rate of the discrete time model of the quantum consensus over a quantum network with $N$ qudits.
Unlike the results obtained for the continuous time model \cite{SaberContQuanArXiv}, the convergence rate of the algorithm depends on the value of $d$ is qudits.

First we expand the density matrix in terms of the generalized Gell-Mann matrices and show that the induced graphs are the Schreier graphs.
Then using the Young Tabloids, we sort the induced graphs obtained from all possible partitions of the integer $N$.
Exploiting the Specht module representation of partitions of $N$, we have shown that the spectrum of the Laplacian corresponding to the less dominant partition in the Hasse Diagram includes that of the one level dominant partition.
Therefore the Laplacian matrix corresponding to partition $(1, 1, \cdots, 1)$ includes the corresponding spectrum of all other partitions.
Based on this result and the Aldous' conjecture \cite{AldousBook}
we have shown that the second smallest eigenvalues $(\lambda_2(\boldsymbol{L}))$ of the Laplacian of all partitions (except ($N$)) in the Hasse diagram are equal.
This is the generalization of the Aldous' conjecture to all partitions (except ($N$)) in the Hasse diagram of integer $N$.
We have shown that the problem of optimizing the convergence rate of the discrete time quantum consensus reduces to optimizing the Second Largest Eigenvalue Modulus (SLEM) of the weight matrix which can be formulated in terms of two eigenvalues, namely, the second smallest $(\lambda_2(\boldsymbol{L}))$ and the greatest $(\lambda_{max}(\boldsymbol{L}))$ eigenvalues of the Laplacian matrix.

Applying the generalization of the Aldous' conjecture, we have proved that $\lambda_2(\boldsymbol{L})$ can be calculated from the Laplacian matrix corresponding to any of the partitions (other than $n = (N)$), where the most suitable one is  partition ($N-1, 1$).
Unlike $\lambda_2(\boldsymbol{L})$, the greatest eigenvalue $(\lambda_{max}(\boldsymbol{L}))$ of the induced graphs corresponding to different partitions are not the same.
Selecting the appropriate induced graph that contains $\lambda_{max}(\boldsymbol{L})$ depends on the value of $N$ and $d$.
For $N \leq d2$, the greatest eigenvalue $(\lambda_{max}(\boldsymbol{L}))$ is obtained from Laplacian matrix corresponding to partition $(1, 1, \cdots, 1)$ which is $2W$, while for larger values of $N$, partition $(1, 1, \cdots, 1)$ is not feasible and the greatest eigenvalue $(\lambda_{max}(\boldsymbol{L}))$ is included in partitions dominant to $(1, 1, \cdots, 1)$.
In the special case of $N = d^2 + 1$, we have provided the semidefinite programming formulation of the problem along with the optimal results.
For values of $N > d^2 + 1$, the problem should be solved per-case.

In the final stage, we have analytically addressed the semidefinite programming formulation of the problem for $N \leq d^2 + 1$ and a wide range of topologies and provided closed-form expressions for the optimal convergence rate and the optimal weights.
For the special case of a quantum network with complete graph topology as its underlying graph, we have included the complete solution of the FDTQC problem for all values of $N$, where group association schemes are employed for obtaining the spectrum of the induced graphs.

The rest of the paper is organized as follows.
Section \ref{sec:Preliminaries} presents some preliminaries including relevant concepts in graph theory, Young tabloids, irreducible representations of finite groups, permutation modules and Specht modules, Cayley and Schreier coset graphs.
The discrete time consensus algorithm and the semidefinite programming formulation of its optimization are presented in Section \ref{sec:DTCPriliminaries}.
Section \ref{sec:ContinuousTimeQuantumConsensus} describes optimization of the discrete time quantum consensus problem and how it can be transformed into optimization of a classical discrete time consensus problem.
In Section \ref{sec:SDP} analytical optimization of the discrete time consensus problem and closed-form expressions for the optimal results for a range of topologies have been presented.
Section \ref{sec:Conclusion}, concludes the paper.

\section{Preliminaries}
\label{sec:Preliminaries}
In this section, we present the fundamental concepts from graph theory, 
irreducible representations of finite groups, Schure's lemma, Permutation Modules, Specht Modules, Cayley and Schreier coset graphs.
Regarding Young tableau, $S_{N}$ group and its representation, the definitions employed here in this paper are adapted from \cite{Segan2001,Gordon2001}.
A comprehensive preliminaries on symmetric group, Young tabloids, Young subgroup and Hasse Diagrams are provided in {\cite[Section~2.2]{SaberContQuanArXiv}}.

%

\subsection{Graph Theory}
\label{sec:GraphTheory}

A graph is defined as $\mathcal{G}= \{ \mathcal{V}, \mathcal{E} \}$ with $\mathcal{V} = \{1, \ldots, N\}$ as the set of vertices and $\mathcal{E}$ as the set of edges.
Each edge $\{i,j\} \in \mathcal{E}$ is an unordered pair of distinct vertices.
If no direction is assigned to the edges, then the graph is called an undirected graph.
Throughout this paper we consider undirected simple graphs with no self-loops
and at most one edge between any two different vertices.
The set of all neighbors of a vertex $i$ is defined as $\mathcal{N}_{i}  \triangleq  \{  j \in \mathcal{V} : \{i,j\} \in \mathcal{E}  \}$.
A weighted graph is a graph where a weight is associated with every edge according to proper map $W: \mathcal{E} \rightarrow \mathbb{R}$, such that if $\{i,j\} \in \mathcal{E}$, then $W(\{i,j\})= \boldsymbol{w}_{ij}$; otherwise $W(\{i,j\}) = 0$.
The edge structure of the weighted graph $\mathcal{G}$ is described through its adjacency matrix $(A_{\mathcal{G}})$.
The adjacency matrix $A_{\mathcal{G}}$ is a $N \times N$ matrix with $\{i,j\}$-th entry $(A_{\mathcal{G}}(i,j))$ defined as below
\begin{equation}
    \nonumber
    \begin{gathered}
        \nonumber  A_\mathcal{G}(i,j) =
	       \begin{cases}
                \boldsymbol{w}_{ij} \quad \text{if} \quad \{i,j\} \in \mathcal{E} \\
                0 \quad \text{Otherwise}
            \end{cases}
     \end{gathered}
\end{equation}
i.e., the $(i,j)-{th}$ entry of $A_{\mathcal{G}}$ is $1$ if vertex $j$ is a neighbor of vertex $i$.
If the graph $\mathcal{G}$ has no self-loops $A_{\mathcal{G}}(i,i) = 0$, i.e., the diagonal elements of the adjacency matrix are all equal to zero.
For undirected graphs the adjacency matrix is symmetric, i.e., $A_{\mathcal{G}}$ is symmetric.
The degree of a vertex $i$ is the sum of the weights on the edges connected to vertex $i$, i.e.
$ d_{i} = \sum_{j=1}^{N} {\boldsymbol{w}_{ij}} $.
The degree matrix $D_{\mathcal{G}}$ of $\mathcal{G}$ is the $N \times N$ diagonal matrix where its $i$-th diagonal element is equal to the degree of vertex $i$ and all non-diagonal elements are equal to zero.
A graph is called connected if there is a path between any two vertices in the graph.
A graph is called a regular graph if all the vertices have the same number of neighbors.
The Laplacian matrix of graph $\mathcal{G}$ is defined as below,
\begin{equation}
    \nonumber
    \begin{gathered}
        \nonumber  L_{\mathcal{G}}(i,j) =
	       \begin{cases}
                D_{\mathcal{G}}(i,i) \quad \text{if} \quad i = j \\
                -A_{\mathcal{G}}(i,j) \quad \text{if} \quad i \neq j
            \end{cases}
     \end{gathered}
\end{equation}
This definition of the Laplacian matrix can be expressed in matrix form as $L_{\mathcal{G}} = D_{\mathcal{G}} - A_{\mathcal{G}}$, where $D_{\mathcal{G}}$ and $A_{\mathcal{G}}$ are the degree and the adjacency matrices of the graph $\mathcal{G}$.
The Laplacian matrix of an undirected graph is a symmetric matrix.
The eigenvalues of the Laplacian matrix $(L_{\mathcal{G}})$ are all nonnegative.
Defining $\bf{1}$ and $\bf{0}$ as vectors of length $N$ with all elements equal to one and zero, respectively, hence for the Laplacian matrix we have $L_{\mathcal{G}} \times \bf{1} = \bf{0}$.
In undirected graphs, the associated Laplacian is a positive semidefinite matrix and its eigenvalues can be arranged in non-decreasing order 
i.e. $0 = \lambda_{1} (L_{\mathcal{G}}) \leq \lambda_{2} (L_{\mathcal{G}}) \leq \cdots \leq \lambda_{N} (L_{\mathcal{G}})$.
The second smallest eigenvalue $\lambda_{2} (L_{\mathcal{G}})$ is known as the algebraic connectivity and reflects the degree of connectivity of the graph \cite{Fiedler1973}.
First introduced in \cite{Fiedler1973}, this eigenvalue is named algebraic connectivity due to its importance in connectivity properties of the graph.
Since then the algebraic connectivity has found applications in analysis of numerous problems including combinatorial optimization problems such as the maximum cut problem, certain flowing process and the traveling salesman problem \cite{BrazilReview2007}.
The algebraic connectivity can be used to define the spectral gap.
The spectral gap gives insight into important properties of the graph such as the mixing time of random walks \cite{Hoory2006}.
In some cases, the term spectral gap is directly used to refer to $\lambda_{2}(L_{\mathcal{G}})$.
A necessary and sufficient condition for the algebraic connectivity to be nonzero, is that the graph $\mathcal{G}$ is connected \cite{Chung1997}.
If the algebraic connectivity of the graph $\mathcal{G}$ is nonzero then $L_{\mathcal{G}}$ is an irreducible matrix i.e. it is not similar to a block upper triangular matrix with two blocks via a permutation \cite{Horn2006}.
The largest eigenvalue $\lambda_{N} (L_{\mathcal{G}})$ of the Laplacian matrix is known as the Laplacian spectral radius of $\mathcal{G}$.

\subsection{Irreducible Representations of Finite Groups }

Here we provide a brief preliminaries on irreducible representations of finite groups.
The topics covered in this subsection hold true for all finite groups in general, but in this paper, we have used them specifically for permutation groups.

For a $d$-dimensional complex vector space $V$, the general linear group $GL(V)$ is the group of all invertible linear transformations of $V$ to itself which is isomorphic to the group of all $d \times d$ invertible complex matrices denoted by $GL(d,\mathbb{C})$.

\begin{definition} {Matrix Representation}
\label{MatrixRepresentation}
\\
A matrix representation of a group $G$ is a group homomorphism
$ G \xrightarrow{\Phi} GL(d,\mathbb{C})$.
Equivalently, to each $g \in G$ is assigned $\Phi(g) \in GL(d,\mathbb{C})$ such that
\newline 1. $\Phi(\epsilon) = I$ the identity matrix, and
\newline 2. $\Phi(gh) = \Phi(g)\Phi(h)$ for all $g, h \in G$.
\newline $d$ is referred to as the dimension of the representation.
From conditions 1 and 2 it can be concluded that $\Phi(g^{-1}) = {\Phi(g)}^{-1}$
\end{definition}

\begin{definition} {Equivalence of Representation}
\label{EquivalenceRepresentation}
\\
Two representations $G \xrightarrow{\Phi} GL(d,\mathbb{C})$ and $G \xrightarrow{\Theta} GL(d,\mathbb{C})$ of $G$ over $\mathbb{C}$, are equivalent if there exist an invertible $d \times d$ matrix $T$ such that for all $g \in G$, the relation
$\Theta(g) = T^{-1} \Phi(g) T$ stands true.
\end{definition}

\begin{definition} {Group Characters}
\label{GroupCharacters}
\\
The trace of the matrix representation $tr \{ \Phi(g) \}$  is referred to as the character of the group and it is denoted by $\chi(g)$.
Equivalent representations have the same character.
\end{definition}

\begin{definition} {$\mathbb{C}G$-module}
\label{CGModule}
\\
The $\mathbb{C}G$-module over vector space $V$ is a group homomorphism
$G \xrightarrow{\Phi} GL(V)$.
Equivalently, $V$ is a $\mathbb{C}G$-module if there is a multiplication, $gv$ of elements of $V$ by elements of $G$ such that
\newline 1. $gv \in V$;
\newline 2. $(hg)v = h(gv)$;
\newline 3. $1v = v$;
\newline 4. $g(cv) = c(gv)$;
\newline 5. $g(u+v) = gv + gu$;
for all $g,h \in G$; $v,u \in V$ and scalars $c \in \mathbb{C}$.
\end{definition}

\begin{definition} {$\mathbb{C}G$ algebra}
\label{CGAlgebra}
\\
By choosing the elements of the finite group $G$ as a basis, one can define the vector space $\mathbb{C}G$ over $\mathbb{C}$ where the elements of group acts naturally on this vector space.
Hence it is a $\mathbb{C}G$-module called regular
module.
Since the vector space $\mathbb{C}G$ is closed under the multiplication of its elements then it forms an algebra called $\mathbb{C}G$-algebra.
\end{definition}

\begin{definition} {Submodule}
\label{Submodule}
\\
A subset $W$ of the $\mathbb{C}G$-module $V$ is said to be a $\mathbb{C}G$-submodule of $V$ if $W$ is a subspace and $wg \in W$ for all $w \in W$ and all $g \in G$.
Hence a $\mathbb{C}G$-submodule of a $\mathbb{C}G$-module $V$ is a subspace which is also an $\mathbb{C}G$-module.
\end{definition}

\begin{definition} {Irreducible $\mathbb{C}G$-module}
\label{IrreducibleModule}
\\
A $\mathbb{C}G$-module $V$ is said to be irreducible if it is non-zero and it has no $\mathbb{C}G$-submodules apart from $\{0\}$ and $V$.
If $V$ has a $\mathbb{C}G$-submodule $W$ which is not equal to $\{0\}$ or $V$, then $V$ is reducible.
Similarly, a representation $G \xrightarrow{\Phi} GL(d,\mathbb{C})$ is irreducible if the corresponding $\mathbb{C}G$-module $\mathbb{C}^{d}$ given by
$gv = \Phi  ( g ) v $   ( where $v \in \mathbb{C}^d, g \in G $)
is irreducible;
and $\Phi$ is reducible if $\mathbb{C}^{d}$ is reducible.
The number of inequivalent irreducible representations of $G$ is equal to the number of conjugacy classes.
\end{definition}

\begin{definition} {Maschke's Theorem}
\label{MaschkeTheorem}
\\
For the finite group $G$ and a nonzero $G$-module $V$ can be written as the direct sum of its irreducible $G$-submodules, i.e. 
$V = W^{(1)} \oplus W^{(2)} \oplus \cdots \oplus W^{(k)}$.
The matrix version of the Maschke's Theorem states that by suitable choice of basis, the matrix representation of the elements of the group $G$ can be made block diagonal, i.e.
\begin{equation}
    \nonumber
    \begin{gathered}
        \Phi(g) = X(g) =
        \left[ \begin{array}{cccc}
            {X^{(1)}(g)} & {0} & {\cdots} & {0}   \\
            {0} & {X^{(2)}(g)} & {\cdots} & {0}   \\
            {\vdots} & {\vdots} & {\ddots} & {\vdots}   \\
            {0} & {0} & {\cdots} & {X^{(k)}(g)}
        \end{array} \right],   \quad \forall g \in G.
    \end{gathered}
\end{equation}
where each $X^{(i)}(g)$ is an irreducible matrix representation of $G$.
\end{definition}

\subsubsection*{Schure's Lemma}
For a given irreducible representation $X$ of $G$
the only matrices $T$ that commute with $X(g)$ for all $g \in G$ are those of the form $T = cI$ i.e., scalar multiples of the identity matrix.

\subsection{Permutation Modules \& Specht Modules }
In this subsection, we introduce concepts from permutation modules and Specht modules.

\begin{definition} {row and column-stabilizer}
\label{stabilizer}
\\
Suppose that the tableau $t$ has rows $R_1, R_2, \ldots, R_l$ and columns $C_1, C_2, \ldots, C_k$. Then
$ R_t = S_{R_1} \times S_{R_2} \times \cdots \times S_{R_l} $
and
$ C_t = S_{C_1} \times S_{C_2} \times \cdots \times S_{C_k} $
are the row-stabilizer and column-stabilizer of $t$, respectively.
\end{definition}

Note that the equivalence classes can be expressed as ${t}=R_t t$.
Given a subset $H \subseteq S_n$, we can form the group algebra sum
$H^{-} = \sum_{\pi \in H}{sgn(\pi)\pi}$.
$sgn(\Pi)$ for every permutation $\Pi\in S_N$ is defined as $sgn(\Pi) = (-1)^{j}$ if $\Pi$ is a product of $j$ transpositions.
For $n=(n_1,n_2,\cdots,n_K)\vdash N$, the Young subgroup of $S_{N}$ corresponding to $n$ is defined as $S_n \overset{def}{=} S_{n_{1}}\times S_{n_{2}}\times\cdots \times S_{n_{K}}$, where $S_{n_{1}}$ permutes $1,2,\cdots,n_1$, $S_{n_{2}}$ permutes $n_1+1, n_1+2, \cdots ,n_1+n_2$ and so on.
The order of the Young subgroup of $n$-shape is $n_1!n_2!\cdots n_K!$.
For column subgroup $C_t$, $k_{t}$ is defined as 
$ k_{t} \stackrel{def}{=} C_{t}^{-} = \sum_{\pi \in C_t}{sgn(\pi)\pi} $.
Considering the definition of $C_t$ and the fact that it is formed from subgroups $S_{C_1}, S_{C_2}, \ldots, S_{C_k}$  then  we have
$k_t$  $=$  $k_{C_1}$ $k_{C_2}$ $\cdots$ $k_{C_k}$.

\begin{definition} {polytabloid}
\label{polytabloidDef}
\\
Given $t$ is a tableau, then the associated polytabloid is $e_t = k_t\{t\}$.
\end{definition}
\begin{lemma} {}
\label{polytabloidLemma}
Let $t$ be a tableau and $\pi$ be a permutation, then $   k_{\pi t} = \pi k_t \pi^{-1}$ and $e_{\pi t} = \pi e_{t}   $.
\end{lemma}
\begin{definition} {Specht module}
\label{SpechtModule}
\\
For any partition $n$, the corresponding Specht module $S^{n}$ is the submodule of $M^{n}$ spanned by the polytabloids $e_t$, where $t$ is of shape $n$.
\end{definition}
The Specht modules corresponding to different partitions are inequivalent and individually $S_{N}$-irreducible representations over the complex field $\mathbb{C}$.
Also they form a complete list of irreducible $S_{n}$ modules over $\mathbb{C}$.
For Specht module $S^{n}$, $S^{n} \otimes S^{(1^n)}$ is also an irreducible $\mathbb{C}$-$S_N$ module isomorphic to its dual or $S^{n^{'}}$ where $n^{'}$ is conjugate of partition $n$ {\cite[Theorem~8.15]{Gordon1978}} i.e. for any given matrix representation of Specht module $X^{(n)} \in S^{n}$
\begin{equation}
    \label{eq:SpechtModuleDuality}
    \begin{gathered}
        X^n(\Pi)sgn(\Pi) \cong X^{n^{'}}(\Pi) \quad \quad \forall \, \Pi \in S_N.
    \end{gathered}
\end{equation}

In the following we explain the Specht module for three important partitions.

1) $n = (N)$: The $e_{(1,2, \ldots, N)} = \boxed{1 \; 2 \; \cdots \; N}$ is the only polytabloid and $S^{(N)}$ is the one-dimensional Specht module corresponding to the trivial representation.
Since $\forall \pi \in S_{N}$, we have $ \pi e_{t} = e_t $.

2) $n = (\underbrace{\scriptstyle 1,1,\ldots,1}_{N}) = (1^N)$: Let $t$ be the standard tableau and $e_t$ as its only polytabloid as below
\begin{spacing}{1.00}
$$ t= \begin{array}{c}     {\boxed{1}}  \\  {\boxed{2}}  \\  {\vdots}  \\  {\boxed{N}}  \end{array},  \quad \quad \quad \quad   e_t = \sum_{\pi \in S_N} {sgn(\pi) \begin{array}{c}     {\boxed{\pi(1)}}  \\  {\boxed{\pi(2)}}  \\  {\vdots}  \\  {\boxed{\pi(N)}}  \end{array} }, $$
\end{spacing}
\vspace{10px}
thus  $\forall \pi \in S_N$ we have $\pi e_{t}  =  sgn (\pi)e_t$.
Hence this is also a one-dimensional or scalar Specht module.

3) $n = (N-1,1)$:
\begin{spacing}{1.00}
$$  e_t = \begin{array}{c}   {\boxed{i \cdots k}}  \\  {\boxed{j} \;\;\;\;\;\;\; }  \end{array}  -  \begin{array}{c}   {\boxed{j \cdots k}}  \\  {\boxed{i} \;\;\;\;\;\;\;\;}  \end{array}  =  j - i  \quad  \text{where} \; j \; \text{is defined as} \quad \{t\}   =   \begin{array}{c}   {\boxed{i \cdots k}}  \\  {\boxed{j}\;\;\;\;\;\;\;}  \end{array}  \stackrel{def}{=}  j $$
\end{spacing}
%
%
Thus 
$  S^{(N-1,1)} = \mathbb{C}[j-i| 1 \leq i < j \leq N] = \{ \sum_{i=1}^{N} {c_i i} | \sum_{i=1}^{N}{c_i} = 0 \}  $.
Note that the dimension of the Specht module in this case is $N-1$.

The permutation module decomposes as
\begin{equation}
    \label{eq:PermutationModuleDecomposition}
    \begin{gathered}
         M^{n^{'}}  =  \bigoplus_{n \unrhd n^{'}} { m_{n, n^{'}}S^{n} }
    \end{gathered}
\end{equation}
with the diagonal multiplicity $m_{n,n} = 1$.
$m_{n,n^{'}}$ is the multiplicity of irreducible representation $S^{n}$ in $M^{n^{'}}$.
Based on the decomposition (\ref{eq:PermutationModuleDecomposition}), the matrix representation of $M^{N}$ is block diagonal where the blocks consists of irreducible Specht representations corresponding to all partitions dominant to $n$.
As an example, $M^{(2,1,1)}  \cong  S^{(2,1,1)} \otimes S^{(2,2)} \otimes 2S^{(3,1)} \otimes S^{(4)} $.
Note that for partition $n^{'}=(1,1,\ldots,1)$, the permutation module $M^{n^{'}}$ is equivalent to regular module and the coefficients $m_{n, n^{'}}$ are equal to the dimension of Specht module corresponding to partition $n$.
Therefore, the regular module includes all irreducible $\mathbb{C}$-$S_N$ modules.

\subsection{Cayley Graph \& Schreier Coset Graph}

Let $\mathcal{H}$ be a group and let $\mathcal{S} \subseteq \mathcal{H}$.
The Cayley graph of $\mathcal{H}$ generated by $\mathcal{S}$ (referred to as the generator set $\mathcal{S}$), denoted by $Cay(\mathcal{H}, \mathcal{S})$, is the directed graph $\mathcal{G} = (\mathcal{V}, \mathcal{E})$ where $\mathcal{V} = \mathcal{H}$ and $\mathcal{E} = \{(x, xs) | x \in \mathcal{H}, s \in \mathcal{S}\}$.
If $\mathcal{S} = \mathcal{S}^{-1}$ (i.e., $\mathcal{S}$ is closed under inverse), then $Cay(\mathcal{H}, \mathcal{S})$ is an undirected graph.
If $\mathcal{H}$ acts transitively on a finite set $\Omega$, we may form a graph with vertex set $\mathcal{V} = \Omega$ and edge set $\mathcal{E} = \{ (\nu, \nu s) | \nu \in \Omega, s \in \mathcal{S} \}$. Similarly, if $\mathcal{Q}$ is a subgroup  in $\mathcal{H}$, we may form a graph whose vertices are the right cosets of $\mathcal{Q}$ , denoted $(\mathcal{H}:\mathcal{Q})$ and whose edges are of the form $\mathcal{E} = \{(\mathcal{Q}h, \mathcal{Q}hs) |\mathcal{Q}h\in (\mathcal{H}:\mathcal{Q}), s \in \mathcal{S}\}$.
These two graphs are the same when $\Omega$ is the coset space $(\mathcal{H}:\mathcal{Q})$, or when $\mathcal{Q}$ is the stabilizer of a point of $\Omega$ and  is called the Schreier coset graph $Sch(\mathcal{H}, \mathcal{S}, \mathcal{Q})$.

\section{Classical Discrete Time Consensus (DTC)}
\label{sec:DTCPriliminaries}

Consider a group of $N$ agents with an undirected underlying connected graph $\mathcal{G} = (\mathcal{V}, \mathcal{E})$,
where each edge $\{i,j\}$ indicate bidirectional communication between agent $i$ and agent $j$.
Let $x_{i}$ be the state of agent $i$.
In the Discrete Time Consensus (DTC) algorithm, each agent's dynamics evolves according to the following state update equation,
\begin{equation}
    \label{eq:ConsensusStateDynamicsEquation}
    \begin{aligned}
        x_i(t+1)  =  \boldsymbol{W}_{i,i} \cdot x_i (t) + \sum_{j \in N_i} { \boldsymbol{W}_{i,j} \cdot x_j(t) } \quad \text{for} \quad i=1, \ldots, N.
    \end{aligned}
\end{equation}
Here $N_i$ is the set of neighbors of vertex $i$.
$\boldsymbol{W}_{i,j}$ is the weight assigned to the edge $(i,j)$.
In other words $\boldsymbol{W}_{i,j}$ is the weight assigned to the state of vertex $j$, when vertex $i$ is updating its state.
The summation of agents' states is preserved in each iteration of (\ref{eq:ConsensusStateDynamicsEquation}), therefore $\boldsymbol{W}_{i,i}  =  1  -  \sum_{j \in N_i} {\boldsymbol{W}_{i,j}}$.
Defining the vector $\boldsymbol{x} = [x_{1}, \cdots, x_{N}]^{T}$ as the vector of states, we can rewrite the above state update formula in compact notation as below,
\begin{equation}
    \label{eq:CTCStateUpdate}
    \begin{aligned}
        \boldsymbol{x}(t+1) = \boldsymbol{W} \times \boldsymbol{x}(t)
    \end{aligned}
\end{equation}
The weight matrix $\boldsymbol{W}$ is a $N \times N$ square matrix with the same sparsity pattern as the adjacency matrix of graph $\mathcal{G}$
and it can be written as $\boldsymbol{W}  =  \boldsymbol{I}  -  \boldsymbol{L}_{\mathcal{G}}$,
in terms of the Laplacian matrix of graph $\mathcal{G}$.
The $(i,j)$-th element of $\boldsymbol{W}$ is the weight assigned to edge $(i,j)$.
Equation (\ref{eq:CTCStateUpdate}) implies that the vector of states at time $t$ is related to their initial state by equation 
$\boldsymbol{x}(t)  =  \boldsymbol{W}^t  \times  \boldsymbol{x}(0)$.
The intention of the distributed average consensus algorithm is to asymptotically compute the average of the initial state of vertices, i.e. $\overline{\boldsymbol{x}}  =  \left( \boldsymbol{1} \cdot \boldsymbol{1}^T / N \right) \times \boldsymbol{x}(0)$.
In other words the weight matrix $\boldsymbol{W}$ is selected so that the 
 relation   $\lim_{t \rightarrow \infty} {\boldsymbol{x}(t)}  =  \lim_{t \rightarrow \infty} {\boldsymbol{W}^t \times \boldsymbol{x}(0)}  =  \frac {  \boldsymbol{1} \times \boldsymbol{1}^T  }  {  N  } \times \boldsymbol{x}(0)$,
holds true, which is possible if the following is satisfied
\begin{equation}
    \label{eq:DTCFinalLimitConstraint900}
    \begin{aligned}
        \lim_{t \rightarrow \infty} {  \boldsymbol{W}^t  }  =  \frac   {  \boldsymbol{1} \times \boldsymbol{1}^T  }    {  N  }.
    \end{aligned}
\end{equation}
Here $\boldsymbol{1}$ is the column vector of all one.

The following are the necessary and sufficient conditions \cite{Xiao04Boyd} for the matrix equation (\ref{eq:DTCFinalLimitConstraint900}) to hold true,
\begin{subequations}
    \label{eq:DCTNecessarySufficient954}
    \begin{gather}
        \boldsymbol{1}^{T} \times \boldsymbol{W} = \boldsymbol{1}^T,  \label{eq:DCTNecessarySufficient954A} \\
        \rho ( \boldsymbol{W} - \boldsymbol{1} \times \boldsymbol{1}^T / N ) < 1.        \label{eq:DCTNecessarySufficient954C}
     \end{gather}
\end{subequations}
Here $\rho(\cdot)$ is the spectral radius of a matrix.
(\ref{eq:DCTNecessarySufficient954A})
means that $\boldsymbol{1}$ is an eigenvector of $\boldsymbol{W}$ corresponding to eigenvalue one, since the weight matrix $\boldsymbol{W}$ is a symmetric matrix.
It can be deduced from (\ref{eq:DCTNecessarySufficient954C}) that all other eigenvalues of $\boldsymbol{W}$ are strictly less than one in magnitude.
It can also be concluded from (\ref{eq:DCTNecessarySufficient954A}) that the sum and the average of vertices' states are preserved in each step of the iteration (\ref{eq:CTCStateUpdate}).
This is due to the fact that $\boldsymbol{1}^T \times \boldsymbol{X}(t+1) = \boldsymbol{1}^T \times \boldsymbol{W} \times \boldsymbol{X}(t) = \boldsymbol{1}^T \times \boldsymbol{X}(t)$.

In \cite{Xiao04Boyd} it has been proved that 
the asymptotic convergence factor can be described in terms of spectral radius as following
\begin{equation}
    \label{eq:DCTConvergenceRate969Assymptotic}
    \begin{gathered}
        r_{asym} (\boldsymbol{W}) = \rho ( \boldsymbol{W} - \boldsymbol{1} \times \boldsymbol{1}^T / N ).
     \end{gathered}
\end{equation}

The  main focus here is on designing the optimal weight matrix for a given network so that it results in the fastest asymptotic convergence rate.
Based on (\ref{eq:DCTConvergenceRate969Assymptotic}), this can be described as the following optimization problem where the objective function is the spectral radius of the matrix $\boldsymbol{W} - \boldsymbol{1} \cdot \boldsymbol{1}^T / N$.
\begin{equation}
    \label{eq:DCTOptimization1005}
    \begin{aligned}
        \min\limits_{\boldsymbol{W}}\quad&\rho(\boldsymbol{W}-\boldsymbol{1}\times\boldsymbol{1}^T/N) \\
        s.t.\quad&\boldsymbol{W}\in\mathscr{S},\quad \boldsymbol{W}=\boldsymbol{W}^T,\quad \boldsymbol{W}\times\boldsymbol{1}=\boldsymbol{1}
    \end{aligned}
\end{equation}
where $\mathscr{S} = \{ \boldsymbol{W} \in \mathbb{R}^{ N \times N } \; | \; \boldsymbol{W}_{i,j}=0 \; \; \textrm{if} \; \; {{i,j}}\not\in\mathcal{E} \; \; \textrm{and} \; \; i \neq j \}$.
The optimization problem (\ref{eq:DCTOptimization1005}) is very similar to the Fastest Markov Chain problem.
The only difference is that in (\ref{eq:DCTOptimization1005}) the weights can have negative values whereas in the Fastest Markov Chain problem, the elements of the transition probability matrix should be non-negative.
If graph $\mathcal{G}$ is a connected graph then there is a feasible $\boldsymbol{W}$ such that $\rho ( \boldsymbol{W} - \boldsymbol{1} \cdot \boldsymbol{1}^T / N ) < 1$.
Note that the optimization problem (\ref{eq:DCTOptimization1005}) is a convex problem since the weight matrix $(W)$ is a symmetric matrix.

For the spectral radius of a square matrix $\boldsymbol{A}\in\mathbb{C}^{N\times N}$ we have $\rho(\boldsymbol{A})=\max\limits_{i}\:(|\lambda_i|)$, where $\lambda_1,...,\lambda_N$ are eigenvalues of the matrix $\boldsymbol{A}$.
Let $\lambda_i(\boldsymbol{W})$ for $i=1,...,N$ be the eigenvalues of the weight matrix $\boldsymbol{W}$ arranged in decreasing order.
$\lambda_1(\boldsymbol{W})=1$ since $\boldsymbol{1}$ is the eigenvector of $\boldsymbol{W}$ corresponding to eigenvalue one.
All eigenvalues of the matrix $\boldsymbol{W}-\boldsymbol{1}\times\boldsymbol{1}^T/N$ are the same as the weight matrix $\boldsymbol{W}$, except the eigenvalue corresponding to eigenvector $\boldsymbol{1}$, which is zero.
Therefore, the spectral radius of $\boldsymbol{W}-\boldsymbol{1}\times\boldsymbol{1}^T/N$ is equal to $\max \{\lambda_2(\boldsymbol{W}),-\lambda_{N}(\boldsymbol{W})\}$.
The term $\max \{\lambda_2(\boldsymbol{W}),-\lambda_{N}(\boldsymbol{W})\}$ is the Second Largest Eigenvalue Modulus (SLEM) of the weight matrix $\boldsymbol{W}$.
Therefore the optimization problem (\ref{eq:DCTOptimization1005}) can be written in terms of the SLEM of the weight matrix as below
\begin{equation}
    \label{eq:DCTOptimization1021}
    \begin{aligned}
        \min\limits_{\boldsymbol{W}}\quad&\max \{\lambda_2(\boldsymbol{W}),-\lambda_{N}(\boldsymbol{W})\} \\
        s.t.\quad&\boldsymbol{W}=\boldsymbol{W}^T,\quad \boldsymbol{W}\times\boldsymbol{1}=\boldsymbol{1}, \quad \forall\{i,j\}\not\in\mathcal{E}\textrm{:} \; \boldsymbol{W}_{i,j}=0.
    \end{aligned}
\end{equation}
Authors in \cite{Xiao04Boyd} have reformulated the optimization problem (\ref{eq:DCTOptimization1021}) as the following semidefinite programming problem,
\begin{equation}
    \label{eq:DCTOptimizationFinal}
    \begin{aligned}
        \min\limits_{s,\boldsymbol{W}}\quad&s \\
        s.t.\quad& -s\boldsymbol{I}\preceq \boldsymbol{W}-\boldsymbol{1}\times\boldsymbol{1}^T/N \preceq s\boldsymbol{I} \\
        &\boldsymbol{W}=\boldsymbol{W}^T,\quad \boldsymbol{W}\times\boldsymbol{1}=\boldsymbol{1}, \quad \forall\{i,j\}\not\in\mathcal{E}\textrm{:} \; \boldsymbol{W}_{i,j}=0.
    \end{aligned}
\end{equation}
$\boldsymbol{I}$ is the identity matrix and $s$ is the scalar optimization variable that bounds the spectral norm of matrix $\boldsymbol{W} - \boldsymbol{1} \times \boldsymbol{1}^T / N$.
$\boldsymbol{A}\preceq \boldsymbol{B}$ means that the matrix $\boldsymbol{B}-\boldsymbol{A}$ is a positive semidefinite matrix.
We refer to problem (\ref{eq:DCTOptimizationFinal}) as the classical Fastest Discrete Time Consensus (FDTC) problem.

\section{Discrete Time Quantum Consensus}
\label{sec:ContinuousTimeQuantumConsensus}

\subsection{Evolution of the Quantum Network}
\label{sec:LindbladMasterEquation}
Considering the quantum network as a composite (or multipartite) quantum system with $N$ qudits, the state space of the quantum network is within the Hilbert space $\mathcal{H}^{\otimes N} = \mathcal{H} \otimes \ldots \otimes \mathcal{H}$, where $\mathcal{H}$ is the d-dimensional Hilbert space over $\mathbb{C}$.
The state of the quantum system is described by its density matrix $(\boldsymbol{\rho})$, which is a positive Hermitian matrix with trace one $(tr(\boldsymbol{\rho}) = 1)$.
An underlying graph $\mathcal{G}=\{ \mathcal{V}, \mathcal{E} \}$ is associated with the quantum network , where $\mathcal{V}=\{1,\ldots, N\}$ is the set of indices for the $N$ qudits, and each element in $\mathcal{E}$ is an unordered pair of two distinct qudits, denoted as $\{j,k\} \in \mathcal{E}$ with $j,k \in \mathcal{V}$.
Permutation group $S_{N}$ acts in a natural way on $\mathcal{V}$ by mapping $\mathcal{V}$ onto itself.
For each permutation $\pi \in S_{N}$ a unitary operator $U_{\pi}$ over $\mathcal{H}^{\otimes N}$ is associated, as 
$U_{\pi} ( Q_{1} \otimes \cdots \otimes Q_{N} ) = Q_{\pi(1)} \otimes \cdots \otimes Q_{\pi(N)}$,
where $Q_{i}$ is an operator in $\mathcal{H}$ for all $i = 1, \ldots, N$.
A special case of permutations is the swapping permutation or transposition where $\pi(j)=k$, $\pi(k)=j$ and $\pi(i) = i$ for all $i \in \mathcal{V}$ and $i \notin {j,k}$.
We denote the swapping permutation between the qudits indices $j$ and $k$ by $\pi_{j,k}$ and the corresponding swapping operator by $U_{j,k}$.
In 
{\cite[Appendix~A]{SaberContQuanArXiv}} the swapping operator $U_{j,k}$ has been expressed as linear combination of the Cartesian product of Gell-Mann matrices.

In \cite{PetersenRef15}, the quantum channels \cite{Nielsen}, \cite{Kraus} are employed as the general framework for studying the evolution of the quantum networks as an open-system.
This is due to the fact that Quantum channels are linear, completely positive (CP) and trace preserving (TP) maps from density operators to density operators, i.e. $\phi: \mathcal{D}(\mathcal{H}^m) \rightarrow \mathcal{D}(\mathcal{H}^m)$, and such maps admit an operator sum representation (OSR) known as Kraus decomposition 
defined as $\phi(\rho) = \sum_{k=1}^{K} { A_k \rho A_{k}^{\dagger} }$,
where $ \sum_{k=1}^{K}{A_{k}^{\dagger} A_{k} }  =  I $ and $K \leq (dim(H))^2$.
Note that this representation is not unique but there is a certain relation between all the possible representations \cite{Nielsen}.
A CPTP map is said unital if $\phi(I) = I$.
A particular set of unital quantum channels is given by random unitaries \cite{Mendl}.
A channel belongs to this class when it admits an OSR with $K$ operators $A_k = \sqrt{p_k} U_k$, with $U_k \in \mathcal{U}(H^m)$ and $p_k \geq 0$ such that $\sum_{k=1}^{K} {p_k} = 1$:
$$	\phi(\rho)  =  \sum_{k=1}^{K} { p_k U_k \rho U_k^{\dagger} }	$$
Such a map can be thought of as a probabilistic mixture of unitary evolutions.

Let $\{ U_i \}_{i=1}^{K}$ be the kraus decomposition of a unital CP map $\phi(.)$ and defining
$\mathcal{A}_{\phi}  =  \{  \rho \in \mathcal{D}(\mathcal{H}^m) | \rho U_i - U_i \rho = 0, \forall i =1, \ldots, K  \}$,
then $ \overline{\rho} \in \mathcal{D}(\mathcal{H}^m) $ is a fixed point of $\phi$, i.e., $\phi(\overline{\rho}) = \overline{\rho}$, if and only if $\overline{\rho} \in A_{\phi}$.

 Let $U_{(j,k)}$ denote the pairwise swap operation of subsystems $(j,k)$  on $\mathcal{H}^{m}$.
If the underlying graph of the network $(\mathcal{G})$ is connected, then the set of fixed points of any CP unital map of the form
\begin{equation}
    \label{eq:StateUpdateMazzarella}
    \begin{gathered}
    	\phi (\rho) = q_0 \rho + \sum_{ (j,k) \in \mathcal{E} } { q_{j,k} U_{(j,k)}^{\dagger} \rho U_{(j,k)} } 	
    \end{gathered}
\end{equation}
with $q_0 + \sum_{(j,k) \in \mathcal{E}} {q_{j,k}} = 1$, $q_0, \{q_{j,k}\} > 0$ coincides with the set of permutation-invariant operators.
Substituting $q_0 = 1 - \sum_{(j,k) \in \mathcal{E}} {q_{j,k}}$ and replacing $q_{j,k}$ with $w_{jk}$ (the positive constant weight over the edge $\{j,k\}$) in (\ref{eq:StateUpdateMazzarella}), we obtain the following state update equation for the quantum network,
\begin{equation}
    \label{eq:Lindblad2}
    \begin{gathered}
        \boldsymbol{\rho(t+1)} =  \boldsymbol{\rho(t)}   +  \sum_{ \{ j , k \} \in \mathcal{E} } { w_{j,k} \left(  U_{jk} \times \boldsymbol{\rho(t)}q
         \times U_{jk}^{\dagger} - \boldsymbol{\rho(t)}  \right) }.
     \end{gathered}
\end{equation}
Note that in order to have the set of transpositions corresponding to the edges of the underlying graph as the generator set $\mathcal{S}$ of the the symmetric group $S_N$, the underlying graph should be connected.

In \cite{PetersenRef15} it is shown that the quantum consensus state of (\ref{eq:Lindblad2}) is as below,
\begin{equation}
    \label{eq:QCMEFinalConsensus}
    \begin{gathered}
        \boldsymbol{\rho}^{*}  =  \frac{1}{N!} \sum_{\pi \in S_{N}} {U_{\pi} \boldsymbol{\rho}(0) U_{\pi}^{\dagger} }.
     \end{gathered}
\end{equation}
and the state update equation reaches this state (i.e. $\lim_{t \rightarrow \infty} {\boldsymbol{\rho}}(t)  =  \boldsymbol{\rho}^{*}$) provided that the underlying graph of the quantum network is connected.

The rest of the analysis presented in this paper focuses on evaluating and optimizing the convergence rate of the state update equation (\ref{eq:Lindblad2}) to its quantum consensus state (\ref{eq:QCMEFinalConsensus}).
In doing so, we expand the density matrix $(\boldsymbol{\rho})$ as the linear combination of the generalized Gell-Mann matrices (introduced in 
{\cite[Appendix~A]{SaberContQuanArXiv}}) as below,
\begin{equation}
    \label{eq:DecompositionDensityGeneral}
    \begin{gathered}
        \boldsymbol{\rho} = \frac{1}{2^N} \sum_{ \mu_{1}, \mu_{2}, \ldots, \mu_{N} = 0 }^{ d^{2} - 1 }  {  \rho_{ \mu_{1}, \mu_{2}, \ldots, \mu_{N} } \cdot \boldsymbol{\lambda}_{\mu_{1}} \otimes  \boldsymbol{\lambda}_{\mu_{2}} \otimes \cdots \otimes \boldsymbol{\lambda}_{\mu_{N}}  }
     \end{gathered}
\end{equation}
where $N$ is the number of particles and $\otimes$ denotes the Cartesian product and $\boldsymbol{\lambda}$ matrices are the generalized Gell-Mann matrices as in 
{\cite[Appendix~A]{SaberContQuanArXiv}}.
Note that due to Hermity of density matrix, its coefficients of expansion $\rho_{ \mu_{1}, \mu_{2}, \ldots, \mu_{N} }$ are real numbers and due to unit trace of $\boldsymbol{\rho}$ we have $\rho_{0,0,\ldots,0} = 1$.

Using the decomposition of $\boldsymbol{\rho}$ in (\ref{eq:DecompositionDensityGeneral}), its permutations can be written as below
\begin{equation}
    \label{eq:DecompositionDensityPermutation}
    \begin{gathered}
        U_{j,k} \times \boldsymbol{\rho} \times U_{j,k}^{\dagger} = \frac{1}{2^N} \sum_{ \mu_{1}, \mu_{2}, \ldots \mu_{N} = 0 }^{ d^{2} - 1 }  {  \rho_{ \mu_{1}, \ldots \mu_{k}, \ldots, \mu_{j}, \ldots, \mu_{N} } \cdot \boldsymbol{\lambda}_{\mu_{1}} \otimes \cdots \boldsymbol{\lambda}_{\mu_{j}} \otimes \cdots \boldsymbol{\lambda}_{\mu_{k}} \otimes \cdots \otimes \boldsymbol{\lambda}_{\mu_{N}}  }
     \end{gathered}
\end{equation}
Note that in (\ref{eq:DecompositionDensityPermutation}) due to permutation operators, the place of indices $\mu_{j}$ and $\mu_{k}$ in the index of parameter $\boldsymbol{\rho}$ are interchanged.
Substituting the density matrix $\boldsymbol{\rho}$ from (\ref{eq:DecompositionDensityGeneral}) and its permutation (\ref{eq:DecompositionDensityPermutation}) in state update equation (\ref{eq:Lindblad2}) and considering the independence of the matrices $\boldsymbol{\lambda}_{\mu_{1}} \otimes \boldsymbol{\lambda}_{\mu_{2}} \otimes \cdots \boldsymbol{\lambda}_{\mu_{N}} $ we can conclude the following for the state update equation (\ref{eq:Lindblad2}),
\begin{equation}
    \label{eq:DensityEquation1}
    \begin{aligned}
        & \rho_{\mu_{1}, \cdots, 
        \mu_{N}}(t+1)  =  \rho_{\mu_{1}, \cdots, 
        \mu_{N}}(t)  +  \\
        &\quad\quad\quad\quad\quad\quad\quad\quad  \sum_{\{j,k\} \in \mathcal{E} }   {  w_{j,k} \left( \rho_{\mu_{1},\cdots,\mu_{k},\cdots,\mu_{j},\cdots,\mu_{N} }(t)  -  \rho_{\mu_{1},\cdots,\mu_{j},\cdots,\mu_{k},\cdots,\mu_{N} }(t)  \right)  } \\
        &\quad\quad\quad\quad\quad\quad\quad\quad\quad\quad\quad\quad\quad\quad\quad\quad\quad  \text{for all     } \mu_{1},\mu_{2},\cdots,\mu_{N}=0,\cdots,d^{2}-1,
    \end{aligned}
\end{equation}

Following the same procedure, the tensor component of the quantum consensus state (\ref{eq:QCMEFinalConsensus}) can be written as below
\begin{equation}
    \label{eq:QuantumConsensusState872}
    \begin{gathered}
        \rho_{\mu_1, \mu_2, \ldots, \mu_N}^{*}  =  \frac{1}{N!} \sum_{\pi \in S_{N}} {\rho_{\pi(\mu_1), \pi(\mu_2), \ldots, \pi(\mu_N)} (0)}
    \end{gathered}
\end{equation}
and for connected underlying graph, the state update equation (\ref{eq:Lindblad2}) reaches quantum consensus, componentwise as below
\begin{equation}
    \nonumber
    \begin{gathered}
        \lim_{t \rightarrow \infty} {    \rho_{\mu_1, \mu_2, \ldots, \mu_N}  (t)    }  =  \rho_{\mu_1, \mu_2, \ldots, \mu_N}^{*},
    \end{gathered}
\end{equation}

Comparing the set of equations in (\ref{eq:DensityEquation1}) with those of the DTC problem in (\ref{eq:CTCStateUpdate}) we can see that the state update equation (\ref{eq:Lindblad2}) is transformed into the classical DTC problem  (\ref{eq:CTCStateUpdate}) with $d^{2N} -1$ tensor component $\rho_{\mu_{1}, \cdots, \mu_{N}}$ as the agents' states. 
Defining $\boldsymbol{X}_Q$ as a column vector of length $d^{2N}$ with components $\rho_{\mu_1, \ldots, \mu_N}$, the state update equation of the classical DTC can be written as below,
\begin{equation}
    \label{eq:QuantumStateUpdate}
    \begin{gathered}
        \boldsymbol{X}_Q(t+1) =\boldsymbol{X}_Q(t) - \boldsymbol{L}_Q \boldsymbol{X}_Q(t).
    \end{gathered}
\end{equation}
$\boldsymbol{L}_Q$ is the corresponding Laplacian matrix defined as 
$\boldsymbol{L}_Q  =  \sum_{\{j,k\} \in \mathcal{E}} {  \boldsymbol{w}_{j,k} (  I_{d^{2N}}-U_{j,k} )  }$,
where $U_{j,k}$ is the swapping operator given in 
{\cite[Appendix~A]{SaberContQuanArXiv}},
provided that $d$ is replaced with $d^2$ which in turn results in Gell-Mann matrices of size  $d^2 \times d^2$.
As explained in section \ref{sec:DTCPriliminaries}, the convergence rate of the obtained DTC problem is dictated by the Second Largest Eigenvalue Modulus (SLEM) of the weight matrix $\boldsymbol{W}_{Q}$ defined as,
\begin{equation}
    \label{eq:FDTQCSLEM}
    \begin{gathered}
        SLEM  =  max\{ \lambda_2( \boldsymbol{W}_{Q} ) , -\lambda_{d^{2N}}( \boldsymbol{W}_{Q} ) \},
    \end{gathered}
\end{equation}
where the weight matrix $(\boldsymbol{W}_{Q})$ is 
$\boldsymbol{W}_{Q}  =  \boldsymbol{I} - \boldsymbol{L}_Q$.
Thus the corresponding Fastest Discrete Time Consensus problem can be written as the following optimization problem,
\begin{equation}
    \label{eq:FCTQCInitial}
    \begin{aligned}
        \max\limits_{s, \boldsymbol{W}_{Q}} \quad &s \\
        s.t. \quad &-s\boldsymbol{I}\preceq \boldsymbol{W}_Q - \boldsymbol{1}\times\boldsymbol{1}^T/N! \preceq s\boldsymbol{I} \\
        &\boldsymbol{W}_Q=\boldsymbol{W}_Q^T,\quad \boldsymbol{W}_Q\times\boldsymbol{1}=\boldsymbol{1}, \quad \forall\{i,j\}\not\in\mathcal{E}\textrm{:} \; \boldsymbol{W}_Q(i,j)=0.
        %
    \end{aligned}
\end{equation}
We refer to this problem as the Fastest Discrete Time Quantum Consensus (FDTQC) problem.

The state update equation (\ref{eq:Lindblad2}) reaches quantum consensus (\ref{eq:QCMEFinalConsensus}), due to the fact that the generating set is selected in a away that the whole group of $S_N$ can be generated and the resultant Cayley graph of $S_N$ is connected.
Even though, the quantum consensus is achieved but surprisingly, the equations in (\ref{eq:DensityEquation1}) indicate that all agents are not able to exchange information with each other.
This is due to the fact that the underlying graph of the DTC problem obtained from (\ref{eq:DensityEquation1}) is not connected and the consensus is not reachable in the same sense as in the classical DTC problem, where the sufficient condition for reaching consensus is the connected underlying communication graph.

From the right hand side of the equation (\ref{eq:DensityEquation1}), we can see that the tensor components $\rho_{\mu_{1}, \cdots, \mu_{N}}$ that can be transformed into each other by permuting their indices are communicating and exchanging information with each other.
These tensor components or agent states correspond to Young tabloids of the same partition which in turn are equivalent to the agents in the classical DTC problem.
Thus the agents belonging to the same partition form the connected components of the underlying graph of the classical DTC problem (\ref{eq:DensityEquation1}).

As mentioned above the underlying graph is a cluster of connected components where each connected graph component corresponds to a given partition of $N$ into $K$ integers, namely $N = n_{1} + n_{2} + \cdots + n_{K}$, where $K \leq d^2$ and $n_{j}$ for $j=1,\ldots,K$  is the number of indices in $\rho_{\mu_1, \mu_2, \ldots, \mu_N}$ with equal values.
For a given partition and its associated Young Tabloids, more than one connected component can be obtained depending on the value of the $\mu$ indices.
As an example consider a quantum network with three qubits and the path graph as its underlying graph.
In this network the values that the $\mu$ indices can take are $0, 1, 2$ and $3$.
For partition $n=(2,1)$ and Young Tabloids $t_n(1,1,2), t_n(1,2,1), t_n(2,1,1)$ and $\mu_1 = 0$  and $\mu_2 = 1$ the obtained underlying graph of the DTC problem is a path graph with three vertices where each vertex corresponds to one of the Young Tabloids mentioned above.
Now for the same partition and Young Tabloids but different values of the $\mu$ indices (e.g. $\mu_1 = 1$  and $\mu_2 = 0$) the obtained underlying graph of the DTC problem is same as that of the previous example.
As a matter of fact for this partition there are $12$ connected components which are identical to each other.
Each one of these connected components has identical impact on the convergence rate of the state update equation (\ref{eq:Lindblad2}) to its quantum consensus state (\ref{eq:QCMEFinalConsensus}).
Therefore for each partition we consider only one of them and we refer to this graph as the induced graph.
The only exception is the case of $N=d^2$ where there is only one connected component corresponding to the partition that all indices are different from each other.
These induced graphs are the same as those noted in \cite{Petersen2015IEEETranAutControl}.

For the given partition $n=(n_1, n_2, \cdots, n_K)$, using the Yamanouchi symbol (introduced in {\cite[Section~2.2]{SaberContQuanArXiv}}) 
a Young tabloid of partition $n$ is uniquely represented by the notation $t_{n}(r_{1}$, $r_{2}$, $\cdots$, $r_{N-1}$, $r_{N})$.
Each Young tabloid $t_{n}(r_{1}$, $r_{2}$, $\cdots$, $r_{N})$ is equivalent to an agent in the induced graph of the DTC problem
and its corresponding coefficient ($\rho_{\mu_{r_{1}}, \mu_{r_{2}}, \cdots, \mu_{r_{N}}}$) is equivalent to the state of that agent.

The DTC equation obtained from (\ref{eq:DensityEquation1}) for partition $n$ is as below,
\begin{equation}
    \label{eq:Interlacing1}
    \begin{aligned}
        &\rho_{   \mu_{r_{1}(m)},  \ldots, \mu_{r_{N}(m)}}(t+1)
        =\rho_{   \mu_{r_{1}(m)},  \ldots, \mu_{r_{N}(m)}(t)} + \quad\quad\quad\quad\quad\quad\quad\quad \\
        &\qquad \qquad \qquad \qquad \qquad \sum_{\{j,l\} \in \mathcal{E} }  {}   w_{j,l} \cdot \left(   \rho_{\mu_{\pi_{j,l} (r_{1}(m))},\ldots, \mu_{\pi_{j,l} (r_{N}(m))}}(t)    \right.
        -  \left.
        \rho_{   \mu_{r_{1}(m)},  \ldots, \mu_{r_{N}(m)} (t)  }
        \right),
    \end{aligned}
\end{equation}
where $m$ varies from $1$ to $\nu = N! / (n_1! \cdot n_2! \cdots n_K!)$
and $\pi_{j,l}$ transposes the $j$-th and $l$-th Yamanouchi symbols i.e. $r_j$ and $r_l$.
Note that for the 
agent states that their
Yamanouchi symbols $(r_{j}, r_{l})$ are equal, the value inside the summation above is zero.

We define the column vector $\boldsymbol{X}_{n}$ as the state vector of the associated DTC problem (\ref{eq:Interlacing1}) of a given partition $n$.
This vector includes the tensor components corresponding to the Young Tabloids of the partition $n$ and it has ${\nu = N!/(n_1! \cdot n_2! \cdots n_K!)}$ elements.
As mentioned above the underlying graph of the DTC problem is a cluster of connected components,
i.e. the weight matrix $\boldsymbol{W}_Q$ is a block diagonal matrix where each block corresponds to one of the connected components,
with  state vector  $\boldsymbol{X}_{n}$.
The state update equation (\ref{eq:QuantumStateUpdate}) for the state vector $\boldsymbol{X}_{n}$ is as below,
\begin{equation}
    \label{eq:StateUpdateEquationXn}
    \begin{gathered}
        \boldsymbol{X}_n(t+1) = \boldsymbol{W}_n \times \boldsymbol{X}_n(t) 
    \end{gathered}
\end{equation}
with $\boldsymbol{W}_n$ as the weight matrix which is one of the blocks in $\boldsymbol{W}_Q$.

The tensor component of the quantum consensus state (\ref{eq:QuantumConsensusState872}) for partition $n$ takes the following form
\begin{equation}
    \label{eq:QuantumConsensusState13}
    \begin{gathered}
        \rho_{   \mu_{r_{1}},  \ldots, \mu_{r_{N}}   }^{*}   =  \frac{1}{N!} \sum_{\pi \in S_{N}} { \rho_{ \mu_{\pi(r_{1})},  \ldots, \mu_{\pi(r_{N})} } }
    \end{gathered}
\end{equation}

As explained in {\cite[Section~2.2]{SaberContQuanArXiv}}, 
$S_N$ acts transitively over the set of Young tabloids or agents and consequently over the set of agent states $(\{  \{ \rho_n (r_1(1)$, $r_2(1)$, $\cdots$, $r_N(1)) \}$,  $\cdots$,  $\{ \rho_n (r_1(\nu)$, $r_2(\nu)$, $\cdots$, $r_N(\nu)) \}  \})$ with the Young subgroup $S_{n}$ as its stabilizer subgroup.
Since the group elements of the Young subgroup do not change the Yamanouchi symbols.
Based on the one to one correspondence between agent states and the right or left cosets of $S_n$ in $S_N$, 
it can be concluded that the connected component is the Schreier coset graph of permutation group $S_{N}$  with Young subgroup $S_{n}$ and generating set consisting of transpositions associated with edges of the underlying graph of the quantum network.
For the case of trivial $S_n$ (i.e. $n=(1, 1, \ldots, 1)$) the Schreier coset graph is reduced to Cayley graph.

In the following we provide some of typical partitions with their corresponding connected graph component.

The most simple case is the one that all indices are the same, i.e. the partition is $n= (n_{1})=(N)$.
The Yamanouchi symbols for this partition are $r_{1} = r_{2} = \cdots = r_{N} = 1$ and the DTC equation (\ref{eq:Interlacing1}) is 
$\rho_{\mu_1, \mu_1, \ldots,\mu_1}(t) = \rho_{\mu_1, \mu_1, \ldots, \mu_1}(0)$, for $\mu_1=0,1,\ldots,d^2-1$.
The induced graph of this partition is the edgeless or the empty graph which is a graph without any edges.
This is the Schreier coset graph $Sch(S_N, \mathcal{S}, S_N)$.
Due to lack of any information exchange between agents, the agent states does not change by time and they maintain their initial values.
Thus for the quantum consensus state (\ref{eq:QuantumConsensusState13}) we have
$\rho_{\mu_1, \mu_1, \ldots,\mu_1}^{*} = \frac{1}{N!} \sum_{\pi \in S_{N}} { \rho_{\pi(\mu_1), \pi(\mu_1), \ldots, \pi(\mu_1)}(0) } = \rho_{\mu_1, \mu_1, \ldots, \mu_1}(0)$,
where the second equality above is based on the fact that agent state $\rho_{\mu_1, \mu_1, \ldots,\mu_1}$ remains intact under the permutation $\pi$ or any exchange of information.

The next nontrivial partition is the case where all $\mu$ indices are the same except for one of them, 
i.e. $n = (N-1, 1)$ and
the Yamanouchi symbols are as $r_{i} = 1$ for $i=\{1,...,N\} \setminus \{j\}$  and $r_{j} = 2$.
Thus the agent state can be written as $\rho_{\mu_1,\ldots,\mu_1,\mu_{2},\mu_1,\ldots,\mu_1}$ where for ease of notation we denote the agent state by the scalar variable $x_{j}$ for $j=1, \ldots, N$.
Hence the DTC equation (\ref{eq:Interlacing1}) for the partition $n=(N-1,1)$  can be written as 
$x_{j}(t+1) = x_{j}(t)+\sum_{k \in \mathcal{N}(j)}  { w_{j,k} ( x_{k}(t) - x_{j}(t) ) }$,
where $\mathcal{N}(j)$ is the set of neighbours of vertex $j$ in the graph $\mathcal{G}$.
Considering $x_{j}$ as the state for vertex $j$, the equation above is same as the classical DTC problem over the underlying graph $\mathcal{G}$
which in turn is the Schreier graph $Sch(S_N, \mathcal{S}, S_{N-1})$.
For this particular partition the induced graph of the partition is same as the underlying graph of quantum network $(\mathcal{G})$.
For the quantum consensus state (\ref{eq:QuantumConsensusState872}) of this partition we have
$\rho_{\mu_1, \mu_1, \ldots,\mu_1, \mu_2}^{*}$ $=$ $\rho_{\mu_1, \mu_1, \ldots,\mu_1, \mu_2, \mu_1}^{*}$ $=$ $\cdots$  $=$ $\frac{1}{N} \sum_{j=1}^{N} { \rho_{\mu_1, \ldots, \mu_1, \underbrace{\scriptstyle\mu_2}_{j\text{-th}}, \mu_1, \ldots, \mu_1}(0) }$
        $=$ $\frac{1}{N} \sum_{j=1}^{N} {x_{j}(0)}$.
Note that in this case the quantum consensus state is same as the final equilibrium state of the classical DTC problem.

For the case that all indices are different, namely for the partition $n = (1,1,\ldots,1)$, the DTC problem is referred to as interchange Process \cite{ProofAldous}.
This case is possible if $N \leq d^2$.
The Yamanouchi symbols for this partition take different values from $1$ to $N$ where no two symbols are equal to each other.
The quantum consensus state (\ref{eq:QuantumConsensusState872}) for this partition is same as (\ref{eq:QuantumConsensusState13}) with the exception that no two $\mu$ indices have the same value.
The induced graph of this partition is the Schreier coset graph $Sch(S_N, \mathcal{S}, e)$ where $e$ is the identity element of $S_N$.
This Schreier coset graph is same as the Cayley graph $(S_N, \mathcal{S})$.

\begin{figure}
  \centering
     \includegraphics[width=100mm]{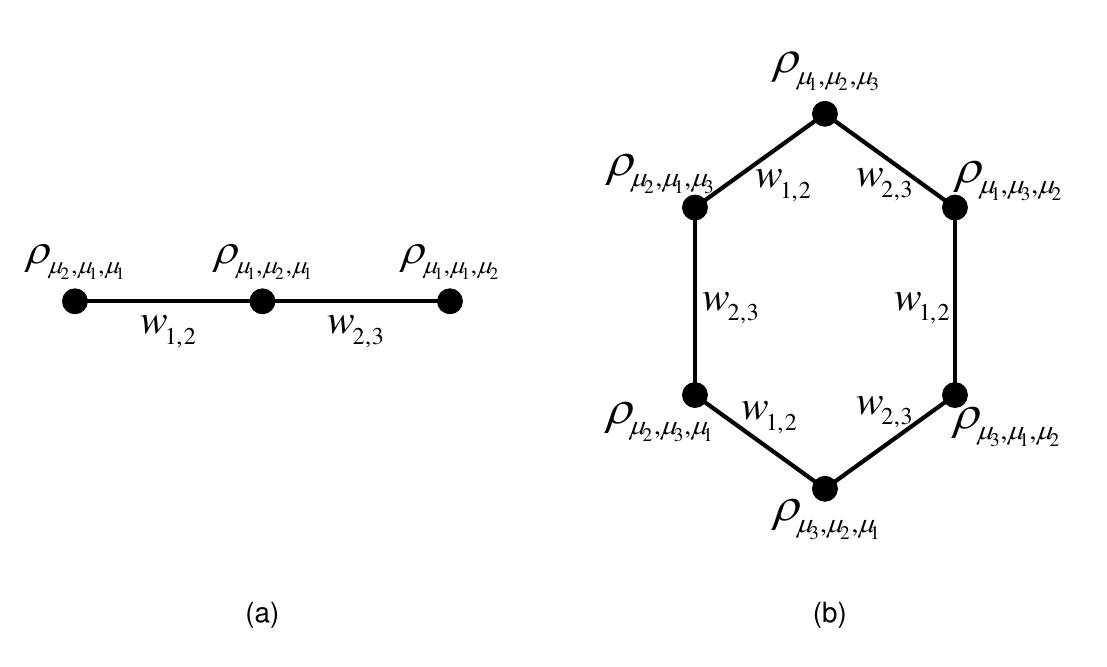}
  \caption{The induced graphs for a path graph with 3 vertices for partitions (a) $n=(2,1)$ and (b) $n=(1,1,1)$.}
  \label{fig:InducedGraphs4Path3}
\end{figure}

As an example, consider the path graph with three vertices (denoted by $\mathcal{G}_{P3}$) as the underlying graph of the quantum network.
For partition $n=(2,1)$ over graph $\mathcal{G}_{P3}$ the induced graph is as depicted in figure \ref{fig:InducedGraphs4Path3} (a) which is same as the underlying graph $\mathcal{G}_{P3}$.
But for partition $n=(1,1,1)$ the induced graph obtained is a cycle graph as depicted in figure \ref{fig:InducedGraphs4Path3} (b).

\subsection{ Spectrum of Induced Graphs }

Based on (\ref{eq:StateUpdateEquationXn}), $\boldsymbol{L}_n$ the Laplacian matrix of the induced graph corresponding to partition $n$ can be written as below,
\begin{equation}
    \label{eq:RobustSolitonDistribution1540}
    \begin{gathered}
        \boldsymbol{L}_n  =  W \cdot \boldsymbol{I}_{\nu} - \boldsymbol{A}_{G(n)}(w),
     \end{gathered}
\end{equation}
Where $W = \sum_{\{i,j\}\in \mathcal{E}}{w_{i,j}}$ and $\boldsymbol{A}_{G(n)}(w) = \sum_{\{i,j\}\in \mathcal{E}} { w_{i,j} \pi_{i,j} }$.
$\boldsymbol{A}_{G(n)}(w)$ is the weighted adjacency matrix of the induced Schreier graph corresponding to partition $n$ and $\pi_{i,j}$ is the matrix representation of transposition of $i \leftrightarrow j$ in the permutation module $M^{(n)}$.
$\boldsymbol{A}_{G(n)}(w)$ is also an element of the $\mathbb{C}$-$S_{N}$ algebra.
According to the decomposition (\ref{eq:PermutationModuleDecomposition}), the matrix  $\boldsymbol{A}_{G(n)}(w)$ is block diagonal where each one of the block matrices denoted by $\tilde{\boldsymbol{A}}_{G(n)}(w)$ consists of matrix representation of $\sum_{\{i,j\}\in \mathcal{E}} { w_{i,j} \pi_{i,j} }$  in irreducible Specht modules of all partitions dominant to $n$.
Therefore $\boldsymbol{L}_n$ is also block diagonal matrix and it can be written as below,
\begin{equation}
    \label{eq:LaplacianPartitionBlockSum}
    \begin{gathered}
        \boldsymbol{L}_n  = diag\{  m_{n,n^{'}} \boldsymbol{B}_{(n^{'})} | \forall n^{'} \unrhd n  \}
        =
        \left[ \begin{array}{cccc}
            {  \boldsymbol{B}_{(N)}  } & {0} & {\cdots} & {0}   \\
            {0}                      & {\boldsymbol{B}_{(N-1,1)}} & {\cdots} & {0}   \\
            {\vdots}                 & {\vdots} & {\ddots} & {\vdots}   \\
            {0}                      & {0} & {\cdots} & {\boldsymbol{B}_{(n)}}
        \end{array} \right],
    \end{gathered}
\end{equation}
where $ \boldsymbol{B}_{(n^{'})} = W \cdot \boldsymbol{I}_{d(n^{'})}  -  \tilde{\boldsymbol{A}}_{G(n^{'})}(w) $ is the representation of Specht module for partition $n^{'}$.
Note that $\boldsymbol{B}_{(N)}$ is a $1 \times 1$ matrix which is equal to zero and it is the only block matrix which has a zero eigenvalue. This is due to the connectivity of the induced Schreier graph.
$\boldsymbol{B}_{(n^{'})}$ is repeated $m_{n,n^{'}}$ times in (\ref{eq:LaplacianPartitionBlockSum}).
For the partition $n=(1,1,\ldots,1)$, the $L_{n}$ in (\ref{eq:LaplacianPartitionBlockSum}) includes all irreducible Specht modules.

From (\ref{eq:SpechtModuleDuality}) for partition $n$ and its conjugate $n^{'}$ we have 
$\tilde{\boldsymbol{A}}_{G(n^{'})}(w)$
        $=$
        $\sum_{\{i,j\} \in \mathcal{E}}$  $ w_{ij}$ $X^{(n^{'})}$ $(\Pi_{ij})  $
        $\cong$
        $\sum_{\{i,j\} \in \mathcal{E}} {  w_{ij} X^{(n)}(\Pi_{ij}) sgn(\Pi_{ij})  }$
        $=$ 
        $-$ $\sum_{\{i,j\} \in \mathcal{E}} {  w_{ij} X^{(n)}(\Pi_{ij})  }$
        $=$
        $-$ $\tilde{\boldsymbol{A}}_{G(n)}(w)$.
This is followed from the fact that the value of function $sgn(\Pi_{i,j})$ in (\ref{eq:SpechtModuleDuality}) for all transpositions $(\Pi_{i,j})$ is $-1$.
Thus the eigenvalues of the matrices $\tilde{\boldsymbol{A}}_{G(n)}(w)$ and  $\tilde{\boldsymbol{A}}_{G(n^{'})}(w)$ are the negative of each other.
As an example, the eigenvalues of the matrix $\tilde{\boldsymbol{A}}_{G(n)}(w)$ corresponding to partitions $n=(N)$ and $n=(1,1,\ldots,1)$ are $W$ and $-W$, respectively.
Note that the matrix $\tilde{\boldsymbol{A}}_{G(n)}(w)$ for both partitions $n=(N)$ and $n=(1,1,\ldots,1)$ is a $1 \times 1$ matrix.
For all other partitions, the eigenvalues of the matrix $\tilde{\boldsymbol{A}}_{G(n)}(w)$  are less than $W$ in absolute value,
since otherwise there would have been two zero eigenvalues and the induced graph would be disconnected.
Thus the eigenvalues $W$ and $-W$ are nondegenerate.

Considering partition $n$ and its one level dominant partition $n^{'}$, from (\ref{eq:LaplacianPartitionBlockSum}) it is obvious that the Laplacian matrix $\boldsymbol{L}_n$ includes all blocks of its one level dominant partition $n^{'}$.
Therefore, the Laplacian corresponding to the less dominant partition $n$ includes the spectrum of the Laplacian corresponding to the dominant partition $(n^{'})$.
Based on this it can be concluded that the Laplacian corresponding to partition $n = (1,1,\ldots,1)$ includes the corresponding spectrum of all other partitions.
In \cite{SaberContQuanArXiv} this has been introduced as the intertwining relation.

Authors in \cite{ProofAldous} have proved that the second smallest eigenvalues $(\lambda_2(\boldsymbol{L}_n))$ (i.e. the spectral gap) of the Laplacian matrices corresponding to partitions $(1,1,\ldots,1)$ and $(N-1, 1)$ (known as the interchange and the random walk processes, respectively) are equal.
This is known as the Aldous' conjecture \cite{AldousBook}.
Considering this result and the intertwining relation described above, it can be concluded that the second eigenvalues $(\lambda_2(\boldsymbol{L}_n))$ of all partitions (except $n=(N)$) in the Hasse diagram are equal to each other.
This is the generalization of the Aldous' conjecture to all partitions (except $(N)$) in the Hasse diagram of $N$.

Thus for optimizing the convergence rate, the second smallest eigenvalue $(\lambda_2(\boldsymbol{L}_Q))$ can be calculated from the Laplacian matrix corresponding to any of the partitions (other than $n=(N)$).
But partition $n=(N-1,1)$ is the most suitable one since its induced graph is identical to the underlying graph of the quantum network and also its corresponding Laplacian matrix $(L_{(N-1,1)})$ is the smallest among the Laplacian matrices of all partitions containing $\lambda_2(\boldsymbol{L}_Q)$.

Unlike $\lambda_2(\boldsymbol{L}_n)$, the greatest eigenvalue $(\lambda_{max}(\boldsymbol{L}_n))$ of the induced graphs corresponding to different partitions are not the same.
Selecting the appropriate induced graph that contains the greatest eigenvalue $(\lambda_{d^{2N}}(\boldsymbol{L}_Q))$ depends on the value of $N$ and $d$, which is explained in the following.

For $N \leq d^2$, all Specht modules including $\boldsymbol{B}_{(1,1,\ldots,1)}$ are present in (\ref{eq:LaplacianPartitionBlockSum}) and therefore, the smallest eigenvalue of $\tilde{\boldsymbol{A}}_{G(1,1,\ldots,1)}(w)$ is $-W$ and the greatest eigenvalue of the laplacian matrix $\boldsymbol{L}_n$ of the induced graph (i.e. $\lambda_{max}(\boldsymbol{L}_n)$) corresponding to partition $n=(1,1,\ldots,1)$ is $2W$.
Thus the semidefinite programming formulation of the FDTQC problem can be written as below,
\begin{equation}
    \label{FDTQCSDPCategory1}
    \begin{aligned}
        \min \quad &s \\ 
        s.t.  \quad \   &\boldsymbol{I} - \boldsymbol{L}_{(N-1,1)} - \boldsymbol{J}_{N}/N  \leq s \cdot \boldsymbol{I}, \\
                             &-s \leq 1-2W
    \end{aligned}
\end{equation}

For $N = d^2 + 1$, the partition $(1,1, \ldots , 1)$ is not feasible and the last matrix block included in (\ref{eq:LaplacianPartitionBlockSum}) is $\boldsymbol{B}_{(2,1^{(N-2)})}$.
Thus $\lambda_{d^{2N}}(\boldsymbol{L}_Q)$ is obtained from the matrix block $\boldsymbol{B}_{(2,1^{(N-2)})}$ corresponding to partition $(2,1^{(N-2)})$.
Based on the relation $ \boldsymbol{B}_{(n)} = W \cdot \boldsymbol{I}_{d(n)}  -  \tilde{\boldsymbol{A}}_{G(n)}(w) $, the largest eigenvalue of $\tilde{\boldsymbol{A}}_{(N-1,1)}(w)$ corresponds to the smallest eigenvalue of $B_{(N-1,1)}$ which is the second smallest eigenvalue $\lambda_2(\boldsymbol{L}_Q)$.
Similarly, the smallest eigenvalue of $\tilde{\boldsymbol{A}}_{(2,1^{(N-2)})}(w)$ corresponds to the largest eigenvalue of $B_{(2,1^{(N-2)})}$   which is the greatest eigenvalue $\lambda_{d^{2N}}(\boldsymbol{L}_Q)$.
On the other hand since partition $(2,1^{(N-2)})$ is the conjugate partition of $(N-1,1)$ then the eigenvalues of $\tilde{\boldsymbol{A}}_{(2,1^{(N-2)})}(w)$ are negative of those of $\tilde{\boldsymbol{A}}_{(N-1,1)}(w)$  and the smallest eigenvalue of $\tilde{\boldsymbol{A}}_{(2,1^{(N-2)})}(w)$ is negative of the largest eigenvalue of $\tilde{\boldsymbol{A}}_{(N-1,1)}(w)$.
Therefore, the eigenvalues $\lambda_2(\boldsymbol{L}_Q)$ and $\lambda_{d^{2N}}(\boldsymbol{L}_Q)$ are the smallest eigenvalue of $W.I - \tilde{\boldsymbol{A}}_{(N-1,1)}(w)$ and the largest eigenvalue of $W.I + \tilde{\boldsymbol{A}}_{(N-1,1)}(w)$, respectively.
Denoting the largest eigenvalue of $\tilde{\boldsymbol{A}}_{(N-1,1)}(w)$ by $Max \, Eig (\tilde{\boldsymbol{A}}_{(N-1,1)}(w))$, the eigenvalues $\lambda_2(\boldsymbol{L}_Q)$ and $\lambda_{d^{2N}}(\boldsymbol{L}_Q)$  are as below,
\begin{equation}
    \label{Lambda2Max1737}
    \begin{gathered}
        \lambda_2(\boldsymbol{L}_Q)  =  W  -  Max \, Eig (\tilde{\boldsymbol{A}}_{(N-1,1)}(w)) \\
        \lambda_{d^{2N}}(\boldsymbol{L}_Q)  =  W  +  Max \, Eig (\tilde{\boldsymbol{A}}_{(N-1,1)}(w)).
    \end{gathered}
\end{equation}
The SLEM (\ref{eq:FDTQCSLEM}) of the FDTQC algorithm is $max \{  1-\lambda_2(\boldsymbol{L}_Q)  ,  |1-\lambda_{d^{2N}}(\boldsymbol{L}_Q)|  \}$, where in the optimal case $1-\lambda_2(\boldsymbol{L}_Q) = \lambda_{d^{2N}}(\boldsymbol{L}_Q) - 1$ and using  (\ref{Lambda2Max1737}) we can conclude that $W = 1$.
Thus the semidefinite programming formulation of the FDTQC problem for $N = d^2 + 1$ can be written as below,
\begin{equation}
    \label{FDTQCSDPCategory2}
    \begin{aligned}
        \min \quad &s \\
        s.t.  \quad \   &\boldsymbol{I} - \boldsymbol{L}_{(N-1,1)} - \boldsymbol{J}_{N}/N  \leq s \cdot \boldsymbol{I}, \\
                             &W = 1
    \end{aligned}
\end{equation}

For values of $N$ larger than $d^2 +1$, the solution to the FDTQC problem should be solved per-case for each value of $N$.
In this paper, we have provided the solution to complete graph topology for all values of $N$.

\section{Optimization of the Second Largest Eigenvalue Modulus (SLEM)}
\label{sec:SDP}

In this section, optimal results for the FDTQC problem over different topologies and values of $N \leq d^2 +1$ is provided.
For complete graph topology we have included the complete solution of the  FDTQC problem  for all values of $N$.

First we provide the optimal weights and the SLEM for all topologies with $N=2,3$ and $4$ vertices which lie in the category of $N \leq d^2$.
Next the optimal results for a range of topologies is reported where the FDTQC problem is solved by linear programming and semidefinite programming methods.
In the final subsection, we provide the complete solution of the  FDTQC problem over complete graph topology for all values of $N$.

For $N \leq d^2$, in the semidefinite programming formulation of the FDTQC problem (\ref{FDTQCSDPCategory1}), the second constraint can be written as $W \leq (1+s)/2$.
Replacing the second constraint with $W \leq (1+s)/2$, semidefinite programming formulation of the FDTQC problem (\ref{FDTQCSDPCategory1}) reduces to the following,
\begin{equation}
    \label{FDTQCSDPCategory1D2}
    \begin{aligned}
        \min \quad &s \\
        s.t.  \quad \   &\boldsymbol{I} - \boldsymbol{L}_{(N-1,1)} - \boldsymbol{J}_{N}/N  \leq s \cdot \boldsymbol{I}, \\
                             &W \leq (1+s)/2.
    \end{aligned}
\end{equation}
This is similar to the semidefinite programming formulation of the Fastest Continuous Time Quantum Consensus (FCTQC) problem in \cite{SaberContQuanArXiv} where the constant $D$ (the upper limit on the total amount of weights) is replaced with $(1+s)/2$.
Thus the same solution procedure as in \cite{SaberContQuanArXiv} can be applied for solving the FDTQC problem (\ref{FDTQCSDPCategory1D2}).

In the same way, for $N = d^2+1$, the semidefinite programming formulation of the FDTQC problem (\ref{FDTQCSDPCategory2}) is similar to the semidefinite programming formulation of the Fastest Continuous Time Quantum Consensus (FCTQC) problem in \cite{SaberContQuanArXiv} where the constant $D$ is replaced with $1$.
Therefore, the same solution procedure as in \cite{SaberContQuanArXiv} can be applied for solving the FDTQC problem (\ref{FDTQCSDPCategory2}) for $N = d^2+1$.

Note that for both cases of $N \leq d^2$ and $N = d^2 + 1$, the optimal value of $\lambda_2(\boldsymbol{L}_{Q})$ (the second smallest eigenvalue of the laplacian matrix $\boldsymbol{L}_{Q}$) is given in \cite{SaberContQuanArXiv} while here we optimize the SLEM $(s)$ of the weight matrix which is $max\{1-\lambda_2(\boldsymbol{L}_{Q}), \lambda_{d^{2N}}(\boldsymbol{L}_{Q}) - 1\}$.
Therefore to use the results in \cite{SaberContQuanArXiv}, the $\lambda_2(\boldsymbol{L}_{Q})$ in \cite{SaberContQuanArXiv} and variable $D$ should be replaced with SLEM $(s)$ and the corresponding values as described above, respectively.

\subsection{Topologies with $N = 2, 3$ and $4$ Vertices}

Here we provide the optimal weights and the SLEM for all possible topologies with $N = 3$ and $4$ vertices which are connected and non-isomorphic.

For a network with $N = 2$ vertices, the only connected and non-isomorphic topology is the path graph with $2$ vertices.
The optimal value of the SLEM for this topology is $0$ and the optimal weight is $1/2$.
For a quantum network with $N=3$ particles, there are two connected topologies, namely path and triangle topologies.
In the path topology with $N=3$ vertices there re two edges which have the same wight.
The optimal value of SLEM and the weights are $3/5$ and $2/5$, respectively.
In triangle topology, there are three edges which have the same weight.
Note that the triangle topology is a complete graph.
The optimal SLEM and weights are $1/3$ and $2/9$, respectively.

\begin{figure}
  \centering
     \includegraphics[width=90mm]{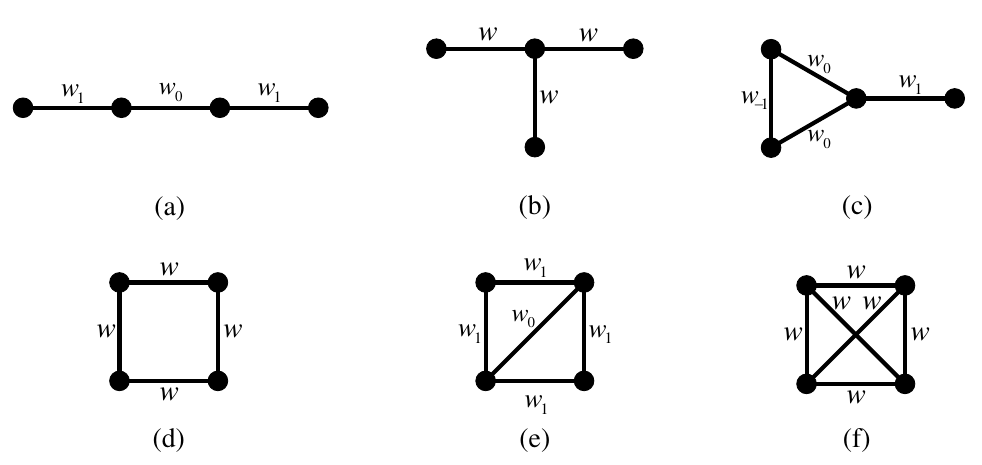}
  \caption{All possible connected underlying topologies with $N=4$ vertices which are non-isomorphic. (a) Path graph, (b) Star graph, (c) Lollipop graph, (d) Cycle graph, (e) Paw graph, (f) Complete graph.}
  \label{fig:N4Graphs}
\end{figure}

In table \ref{tab:N4SLEMOptimal}, all possible connected topologies with $N=4$ vertices (as depicted in figure \ref{fig:N4Graphs}) are listed along with their optimal SLEM and weights.
Note that this case lies in he category of $N \leq 4$.
\begin{table*}
    \label{tab:N4SLEMOptimal}
    \centering
    \caption {All non-isomorphic connected topologies with $N=4$ vertices along with their optimal SLEM and weights}{
    \begin{tabular}{|c|c|c|} \hline
    {Topology}                          & {SLEM}                                            & {Weights}   \\ \hline
    {\multirow{2}{*}{Path}}         & {\multirow{2}{*}{$9/11$}}                         & {$w_0 = 4/11$}  \\
    {}                              & {}                                                & {$w_1 = 3/11$}  \\ \hline
    {Star}                          & {$5/7$}                                           & {$w = 2/7$}  \\ \hline
    {\multirow{3}{*}{Lollipop}}     & {\multirow{3}{*}{${(6+\sqrt{3})}/{11}$}}            & {$w_{-1}=(9-4\sqrt{3})/66$}  \\
    {}                              & {}                                                & {$w_{0}=(6+\sqrt{3})/33$}  \\
    {}                              & {}                                                & {$w_{1}=(6+\sqrt{3})/22$}  \\  \hline
    {Cycle}                         & {$3/5$}                                           & {$w = 1/5$}  \\ \hline
    {\multirow{2}{*}{Paw Graph}}    & {\multirow{2}{*}{$3/5$}}                          & {$w_{0}=0$}  \\
    {}                              & {}                                                & {$w_{0}=1/5$}  \\ \hline
    {Complete Graph}                & {$1/2$}                                           & {$w = 1/8$}  \\ \hline
    \end{tabular}}
\end{table*}

\subsection{Optimization of the SLEM using Linear Programming}

In this subsection, we provide the optimal weights and the SLEM for a number of topologies where linear programming is employed for solving the FDTQC problem \cite{BoydBook2004}.
The optimal results are listed in Table \ref{tab:LinearProgrammingSLEMOptimal}.

\begin{table*}
    \label{tab:LinearProgrammingSLEMOptimal}
    \centering
    \caption {The optimal SLEM and weights of topologies that FDTQC problem is solved using linear programming.}{
    \begin{tabular}{|c|c|c|} \hline
    {Topology}                              & {For $N \leq d^2$}                                    & {For $N = d^2 + 1$}   \\ \hline
    {\multirow{2}{*}{Complete Graph}}       & {$SLEM = (N-2)/N$}                                    & {$SLEM = (N-3)/(N-1)$}  \\
    {}                                      & {$w = 2/N^2$}                                         & {$w = 2/(N^2-N)$}  \\ \hline
    {\multirow{2}{*}{Cycle}}                & {$SLEM=\frac{N-1+cos(2\pi/N)}{N+1-cos(2\pi/N)}$}      & {$SLEM=\frac{N - 2(1-cos(2\pi/N)}{N}$}  \\
    {}                                      & {$w = 1 / ( N + 1 - cos(2\pi/N) )$}                   & {$w = 1 / N$}  \\ \hline
    {\multirow{2}{*}{Simple Star}}          & {$SLEM=(2N-3)/(2N-1)$}                                & {$SLEM  =  (N-2) / (N-1)$}  \\
    {}                                      & {$w = 2 / (2N-1)$}                                    & {$w = 1 / (N-1)$}  \\ \hline
    {\multirow{2}{*}{CPETG}}                & {$SLEM=(2\alpha-1)/(2\alpha+1)$}                      & {$SLEM  =  (\alpha-1) / \alpha$}  \\
    {}                                      & {$w_i=2/((1+2\alpha)\lambda_{i,2})$}                  & {$w_i=1/(\alpha \cdot \lambda_{i,2} )$}  \\ \hline
    {\multirow{2}{*}{Prism}}                & {$SLEM  =  \frac{ 2 N_1 N_2 - N_1 - N_2 - 1 } { 2 N_1 N_2 - N_1 - N_2 + 1 }$}                                                  & {$SLEM  =  \frac{ 2 N_1 N_2 - N_1 - N_2 - 2 } { 2 N_1 N_2 - N_1 - N_2 }$}  \\
    {}                                      & {$w_i = \frac{2} {  (2 N_1 N_2 - N_1 - N_2 + 1)N_i }$}                                                  & {$w_i = \frac{2} {  (2 N_1 N_2 - N_1 - N_2)N_i } $}  \\ \hline
    \end{tabular}}
\end{table*}

First topology reported in Table \ref{tab:LinearProgrammingSLEMOptimal} is the complete graph topology where each vertex is connected to every other vertex in the graph.
Due to the symmetry of complete graph, all edges in the graph have the same weight (denoted by $w$).
Second topology in Table \ref{tab:LinearProgrammingSLEMOptimal} is the Cycle graph which is edge transitive and therefore the optimal weight on all edges is the same.
In simple star topology $N-1$ vertices are connected to one central vertex.
The Cartesian Product of Edge Transitive Graphs (CPETG) is the result of Cartesian product of $m$ edge-transitive weighted graphs.
The structure of the Laplacian matrix for this topology is described in \cite{SaberContQuanArXiv}.
$\lambda_{i,2}$ is the second smallest eigenvalue of the unweighted Laplacian of the $i$-th edge-transitive matrix and $\alpha$ is $\alpha = \sum_{i=1}^{m} {  (\tilde{N}_i \cdot E_i ) / ( N_i \cdot \lambda_{i,2} )  }$, where $\tilde{N}_i = \prod_{j=1}^{i}{N_j}$ and $N_i$ and $E_i$ are the number of vertices and edges in the $i$-th edge-transitive graph, respectively.
The Prism topology is the Cartesian product of two complete graphs, each with $N_1$ and $N_2$ vertices which is a special case of the CPETG topology.

\subsection{Optimization of the SLEM using Semidefinite Programming}

In this subsection, we provide the optimal results for a number of topologies where the FDTQC problem can be solved using Semidefinite Programming \cite{BoydBook2004}.
The topologies included in this subsection are Complete Core Symmetric (CCS) star, CCS star with two types of branches, Symmetric star, Palm, two Coupled Complete Graphs and Lollipop topologies.
The results provided here are 
presented in Tables \ref{tab:SDPSLEMOptimalEqual} and \ref{tab:SDPSLEMOptimalGreater} for $N \leq d^2$ and $N = d^2 + 1$, respectively.

%
%
\begin{figure}
  \centering
     \includegraphics[width=120mm]{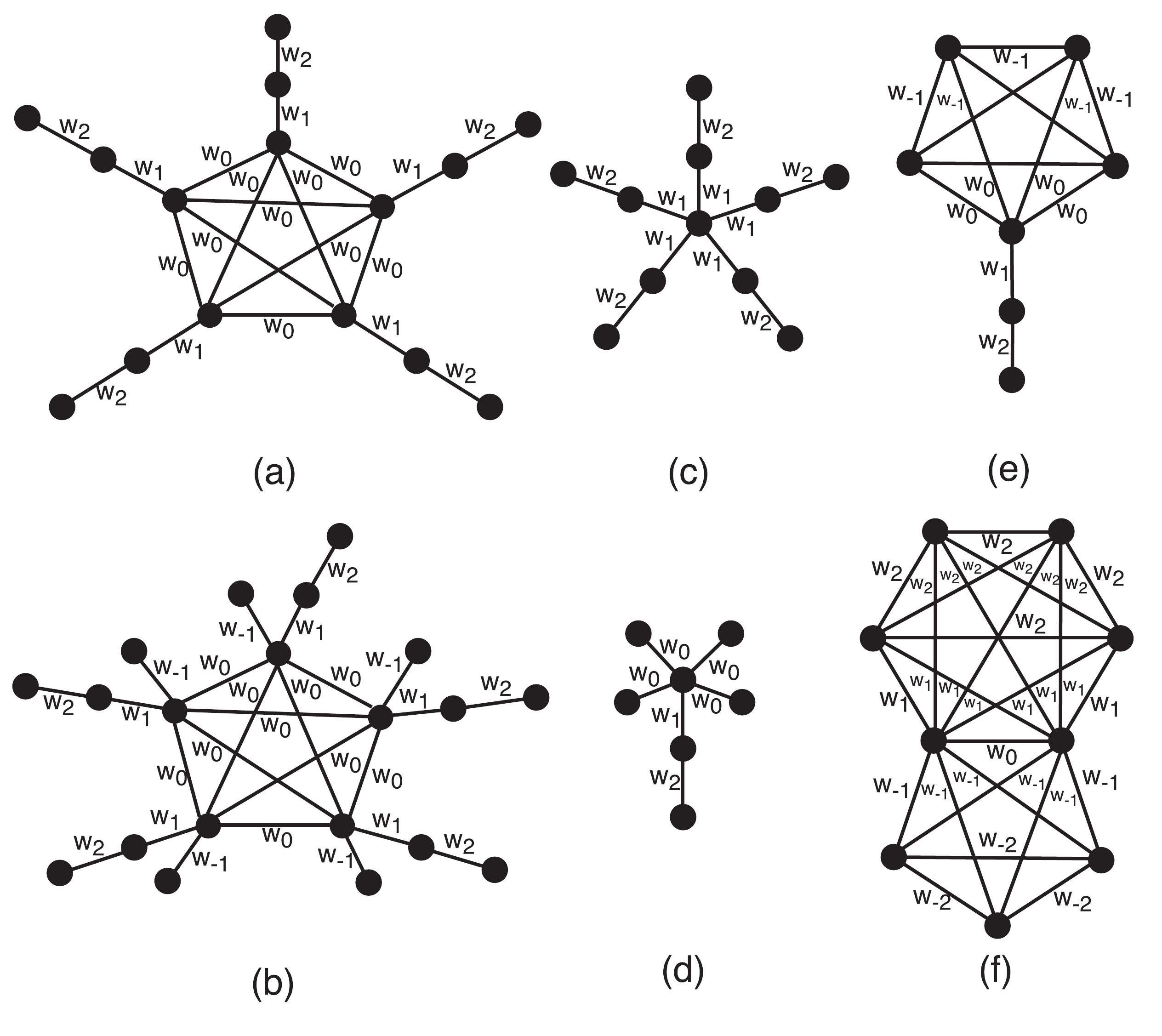}
  \caption{(a) Weighted Complete-Cored Symmetric star topology with $p=5$ branches of length $q=2$. (b) The weighted graph of CCS star graph with two types of branches with $p=5$ branches of length $q_1=1$ and $q_2=2$. (c) A symmetric star graph with $p=5$ branches of length $q=2$. (d) A palm graph with parameters $p=4$ and $q=2$. (e) The weighted Lollipop graph with parameters $p=4$ and $q=2$. (f) The two coupled complete graphs topology with parameters $N_1 = 3$, $N_2 = 2$ and $N_3 = 4$.} 
  \label{fig:SDPGraphs}
\end{figure}

\begin{table*}
    \label{tab:SDPSLEMOptimalEqual}
    \centering
    \caption {The optimal SLEM and weights of topologies that FDTQC problem is solved using Semidefinite programming for $N \leq d^2$.}{
    \begin{tabular}{|c|c|} \hline
    {Topology}                              & {Optimal results for $N \leq d^2$}                                    \\ \hline
    {\multirow{3}{*}{CCS Star}}             & {$SLEM = \frac{\alpha - 3}  {\alpha + 3}$}                                      \\
    {}                                      & {$w_0  =  \frac { 3(2p-2+q\sqrt{2p(p-1)}) }    {  p(p-1)(3p-3+3q\sqrt{2p(p-1)} + 2pq^2 +pq)  }  \times \frac{\alpha}{\alpha+3}$}                                           \\
    {}                                      & {$w_j  =  \frac  {  3 \left(   \sqrt{2p(p-1)}(q-j+1) + p(q-j+1)(q+j)   \right)  }  {  (q+1) \left(3p \left( p-1+q\sqrt{2p(p-1)} \right) + p^2 q(2q+1)\right)  }  \times \frac{ \alpha } { \alpha + 3 }$ for $j\in[1,q]$}                                           \\ \hline
%
%
    {\multirow{4}{*}{\begin{minipage}{1.70in}CCS Star  with $p=2$\end{minipage}}}                & {$SLEM =  \frac {(q+1)(2q+1)(2q+3) - 3}  {(q+1)(2q+1)(2q+3) + 3}$}        \\
    {}                                      & {$w_0 = \frac{3(q+1)^2}{3+(2q+3)(2q+1)(q+1)}$}                     \\
    {}                                      & {$w_j  =  \frac 	{   3  \left(  (q+1)^2 - j^2  \right)	}     {	(q+1)(2q+1)(2q+3) + 3	}$  for $j \in [1,q]$}                     \\  \hline
    {\multirow{4}{*}{\begin{minipage}{1.70in}CCS Star with two types of branches\end{minipage}}}                & {$SLEM = \frac{A - 3}{A + 3}$}        \\
    {}                                      & {$w_0 = \frac{6 }{A+3}\times \frac { 2(p-1)(q_1+q_2+1) + \sqrt{2p(p-1)}(q_1(q_1+1)+q_2(q_2+1)) }  {2p(p-1)}$}                     \\
    {}                                      & {$w_j  =  \frac{6 }{A+3} \times \frac {\sqrt{2p(p-1)}(q_1+j+1) + p(q_1+j+1)(q_1-j )} {2p}$   for  $j=[-q_1,-1]$}                     \\
    {}                                      & {$w_j  =  \frac{6 }{A+3} \times \frac {\sqrt{2p(p-1)}(q_2-j+1) + p(q_2-j+1)(q_2+j )} {2p}$   for  $j=[1,q_2]$}                     \\  \hline
    {\multirow{2}{*}{\begin{minipage}{1.70in}Symmetric Star\end{minipage}}}                & {$SLEM = \frac{ pq(q+1)(2q+1) - 3 }{ pq(q+1)(2q+1) + 3 }$}        \\
    {}                                      & {$w_j = \frac{ 3 (q+j) (q-j+1) }{ pq(q+1)(2q+1) + 3 }$, for $j \in [1,q]$}            \\  \hline
    {\multirow{3}{*}{\begin{minipage}{1.70in}Palm with $2p > q(q + 1)$\end{minipage}}}                & {$SLEM = \frac   {   6p+q(q+1)(2q+1) - 3   }      {   6p+q(q+1)(2q+1) + 3   }$}        \\
    {}                                      & {$w_0 = \frac   {   6   }      {   6p + q(q+1)(2q+1) + 3   }$}                     \\
    {}                                      & {$w_j = \frac   {   (q-j+1) \left(  (q+1)(2q+1) + p(q+j)  \right)   }      {   2(p+q+1)   }$ for $j \in [1,q]$}                     \\  \hline
    {\multirow{3}{*}{\begin{minipage}{1.70in}Palm with $2p \leq q(q + 1)$\end{minipage}}}                & {$SLEM = \frac   {   (q+1)(q+2) \left( 6+q(q+4p+1) \right)   -   6(p+q+1)   }      {   (q+1)(q+2) \left( 6+q(q+4p+1) \right)   +   6(p+q+1)   }$}        \\
    {}                                      & {$w_0 = \frac   {   6(q+1)(q+2)   }      {   (q+1)(q+2) \left( 6+q(q+4p+1) \right)   +   6(p+q+1)   }$}                     \\
    {}                                      & {$w_j = \frac   {   6(q-j+1)    \left(  p(q+j+2) + (q+1)j  \right)   }      {   (q+1)(q+2) \left( 6+q(q+4p+1) \right)   +   6(p+q+1)   }$ for $j \in [1,q]$}                     \\  \hline
    {\multirow{4}{*}{\begin{minipage}{1.70in}Lollipop with $\qquad\qquad\quad$ $\sqrt{2p(p+1)} \geq q(q + 1)$\end{minipage}}}                & {$SLEM = \frac   {   A - 6(p+q+1)   }      {   A + 6(p+q+1)   }$}        \\
    {}                                      & {$w_0 = \frac   {   6(q+1) \left( 2(p+1) + q\sqrt{2p(p+1)} \right)    }    {   A + 6(p+q+1)   }$}                     \\
    {}                                      & {$w_{-1} = \frac   {   12(p+q+1) - 6(q+1) \left( 2(p+1) + q\sqrt{2p(p+1)} \right)    }    {   p \left( A + 6(p+q+1) \right)   }$}                     \\
    {}                                      & {$w_{j} = \frac   {   6(q-j+1) \left( \sqrt{2p(p+1)} + p(q+j) + q+1 \right)    }    {   p \left( A + 6(p+q+1) \right)   }$   for   $j=[1,q]$}                     \\  \hline
    \end{tabular}}
\end{table*}

\begin{table*}
    \label{tab:SDPSLEMOptimalGreater}
    \centering
    \caption {The optimal SLEM and weights of topologies that FDTQC problem is solved using Semidefinite programming for $N = d^2+1$.}{
    \begin{tabular}{|c|c|} \hline
    {Topology}                              & {Optimal results for $N = d^2 + 1$}                                    \\ \hline
    {\multirow{3}{*}{CCS Star}}             & {$SLEM = \frac {\alpha - 6} {\alpha}$}                                      \\
    {}                                      & {$w_0  =  \frac 	{	3  \left(  2p - 2 + q\sqrt{2p(p-1)}  \right) 	} 	{	p(p-1) \left(  3p-3+3q\sqrt{2p(p-1)} + 2pq^2 + pq  \right)	}$}                                           \\
    {}                                      & {$w_j  =  \frac 	{   3  \left(  \sqrt{2p(p-1)}(q-j+1) + p(q-j+1)(q+j)  \right)	}     {	3p(q+1) \left(  p-1+q\sqrt{2p(p-1)}  \right)  + p^2 q (q+1) (2q+1)	}$   for   $j \in [1,q]$}                                           \\ \hline
    {\multirow{4}{*}{\begin{minipage}{1.70in}CCS Star \\ with $p=2$\end{minipage}}}                & {$SLEM = 1 - \frac {6} {(q+1)(2q+1)(2q+3)}$}        \\
    {}                                      & {$w_0  =  \frac 	{	3(q+1) 	} 	{	(2q+3)(2q+1)	}$}                     \\
    {}                                      & {$w_j  =  \frac 	{   3  \left(  (q+1)^2 - j^2  \right) } { (q+1)(2q+1)(2q+3) }$  for $j \in [1,q]$}                     \\  \hline
    {\multirow{4}{*}{\begin{minipage}{1.70in}CCS Star with two types of branches\end{minipage}}}                & {$SLEM = 1  -  \frac{ 6 }{ A }$}        \\
    {}                                      & {$w_0 = \frac{6 }{A} \times \frac { 2(p-1)(q_1+q_2+1) + \sqrt{2p(p-1)}(q_1(q_1+1)+q_2(q_2+1)) }  {2p(p-1)}$}                     \\
    {}                                      & {$w_j  =  \frac{6 }{A} \times \frac {\sqrt{2p(p-1)}(q_1+j+1) + p(q_1+j+1)(q_1-j )} {2p}$ for $j\in[-q_1,-1]$}                     \\
    {}                                      & {$w_j  =  \frac{6 }{A} \times \frac {\sqrt{2p(p-1)}(q_2-j+1) + p(q_2-j+1)(q_2+j )} {2p}$ for $j\in[1,q_2]$}                     \\  \hline
    {\multirow{2}{*}{\begin{minipage}{1.70in}Symmetric Star\end{minipage}}}      & {$SLEM = 1  -  \frac{ 6 }{ pq(q+1)(2q+1) }$}        \\
    {}                                      & {$w_j = \frac{ 3 (q+j) (q-j+1) }{ pq(q+1)(2q+1) }$ for $j\in [1,q]$}                     \\  \hline
    {\multirow{3}{*}{\begin{minipage}{1.70in}Palm with $2p > q(q + 1)$\end{minipage}}}                & {$SLEM = \frac   {   6p + q(q+1)(2q+1)   -   6   }      {   6p + q(q+1)(2q+1)   }$}        \\
    {}                                      & {$w_0 = \frac   {   6   }      {   6p + q(q+1)(2q+1)   }$}                     \\
    {}                                      & {$w_j = \frac   {   (q-j+1) \left(  (q+1)(2q+1) + p(q+j)  \right)   }      {   2(p+q+1)   }$  for  $j \in [1,q]$}                     \\  \hline
    {\multirow{3}{*}{\begin{minipage}{1.70in}Palm with $2p \leq q(q + 1)$\end{minipage}}}                & {$SLEM = 1  -  \frac   {   12(p+q+1)   }      {   (q+1)(q+2)  \left(  6+q(q+4p+1)  \right)   }$}        \\
    {}                                      & {$w_0 = \frac   {   6(q+1)(q+2)   }      {   (q+1)(q+2)  \left(  6+q(q+4p+1)  \right)   }$}                     \\
    {}                                      & {$w_j = \frac   {   6(q-j+1) \left(  p(q+j+2) + (q+1)j  \right)   }      {   (q+1)(q+2) \left(  6 + q(q+4p+1)  \right)   }$  for  $j \in [1,q]$}                     \\  \hline
    {\multirow{4}{*}{\begin{minipage}{1.70in}Lollipop with $\qquad\qquad\quad$ $\sqrt{2p(p+1)} \geq q(q + 1)$\end{minipage}}}                & {$SLEM = 1 - \frac   {   12(p+q+1)   }      {   A   }$}        \\
    {}                                      & {$w_0 = \frac   {   6(q+1) \left(  2(p+1) + q \sqrt{2p(p+1)}  \right)   }      {   A (p+1)  }$}                     \\
    {}                                      & {$w_{-1} = \frac   {   (p+1)A  -  12(p+1)(p+q+1) - 6(q+1) \left(  2(p+1) + q \sqrt{2p(p+1)}  \right)   }      {   p(p+1)A  }$}                     \\
    {}                                      & {$w_{j} = \frac   {   6(q-j+1) \left( \sqrt{2p(p+1)} + p(q+j) + q+1 \right)    }    {   A    }$  for  $j \in [1,q]$}                     \\  \hline
    \end{tabular}}
\end{table*}

Complete-Cored Symmetric (CCS) star topology with parameters $(p,q)$ consists of $p$ path branches of length $q$ (each with $q$ vertices), where branches are connected to each other at one end to form a complete graph in the core.
A CCS star graph with parameters $p=5$, $q=2$ is depicted in figure \ref{fig:SDPGraphs}(a).
Automorphism of the CCS star graph and the structure of its Laplacian matrix are discussed in detail in \cite{SaberContQuanArXiv}.
In the results presented in Tables \ref{tab:SDPSLEMOptimalEqual} and \ref{tab:SDPSLEMOptimalGreater} for CCS star topology, variable $\alpha$ is 
$3(p-1)(q+1) + 3\sqrt{2p(p-1)}q(q+1) + pq(q+1)(2q+1)$.
In the special case of the CCS star topology with $p=2$, this topology reduces to the path topology with even number of vertices.
CCS Star topology with two types of branches has three parameters $(p, q_1, q_2)$, where $p$ is the number of branches from each type and $q_1$ and $q_2$ are the length (number of edges) of branches of first type and second type, respectively.
A CCS star graph with two types of branches with parameters $p = 5$, $q_1 = 1$ and $q_2 = 2$ is depicted in figure \ref{fig:SDPGraphs}(b).
In the results presented in Tables \ref{tab:SDPSLEMOptimalEqual} and \ref{tab:SDPSLEMOptimalGreater} for CCS star topology with two types of branches, variable $A$ is equal to 
$3(p-1)(q_1 + q_2 +1)  +  q_1 ( q_1 + 1 ) \left(  p(2q_1+1 + 3\sqrt{2p(p-1)} )  \right)   +   q_2(q_2+1)  \left(  p(2q_2+1) + 3\sqrt{2p(p-1)}  \right)$.
The symmetric star topology (as depicted in figure \ref{fig:SDPGraphs}(c) with parameters $p = 5$ and $q = 2$) is formed from connecting $p$ path branches (each with $q$ vertices) to a central vertex.
The palm topology is a path graph with $q$ vertices connected to the central vertex of a star graph with $p$ branches as shown in figure \ref{fig:SDPGraphs}(d) for parameters $p=4$ and $q=2$.
Lollipop topology is a path graph (with $q$ vertices) connected to one of the vertices in a complete graph with $p+1$ vertices.
A Lollipop topology with parameters $p=4$ and $q=2$ is depicted in figure \ref{fig:SDPGraphs}(e) along with the weights assigned to the edges.
In the results presented in Tables \ref{tab:SDPSLEMOptimalEqual} and \ref{tab:SDPSLEMOptimalGreater} for Lollipop topology, variable $A$ is equal to 
$6(p-1)(p+q+1)  + (q+1) \left(  6q\sqrt{2p(p+1)} + 6(p+1) + pq(2q+1) + q(q^2-1)  \right)$.
In the case of the Lollipop topology with $\sqrt{2p(p+1)} < q(q + 1)$ the optimal value of $w_{-1}$ is zero and the topology reduces to the Palm topology.

In two coupled complete graphs topology, two complete graphs each with $N_1 + N_2$ and $N_2 + N_3$ vertices respectively, share $N_2$ vertices.
In figure \ref{fig:SDPGraphs}(f) two coupled complete graphs with parameters $N_1 = 3$, $N_2 = 2$ and $N_3 = 4$ is depicted.
Due to the symmetry of the complete graphs weights can be divided into five groups.
$w_{-2}$ is the weight on edges connecting the $N_1$ vertices on the left complete graph to each other and $w_{-1}$ is the weight on the edges connecting the $N_1$ vertices on the left complete graph to the $N_2$ vertices in the middle.
$w_0$ is the weight on edges connecting the $N_2$ vertices in the middle to each other.
Similarly the weights $w_1$ and $w_2$ are defined for the weights on the edges of the complete graph on the right-hand side of the topology.

\begin{itemize}
  \item For $N \leq d^2 $ and symmetric case  $N_1 = N_3$, if $N_1 < N_2/2$ the optimal SLEM is $       ( 4 N_1 N_2 + (N_2 - 1)(N_2 - 2N_1)  - N_2 ) / (   4 N_1 N_2 + (N_2 - 1)(N_2 - 2N_1) + N_2 )$ and the optimal weights are $w_1 = w_{-1} = {2}/(  4 N_1 N_2 + (N_2 - 1)(N_2 - 2N_1) + N_2 )$,   $w_2 = w_{-2} = 0$ and $w_0 = ( N_2 - 2N_1 ) / N_{2}^{2}$.
\item For $N \leq d^2 $ and symmetric case  $N_1 = N_3$, if $N_1 \geq N_2/2$ the optimal SLEM is equal to $(4 N_{1}-1) / (4 N_{1}+1)$ and the optimal weights are 
$w_2 = w_{-2} = 0$, $w_1 = w_{-1} = {2} / (N_{2}(4 N_{1}+1))$ and $w_0 = 0$.
\item For $N = d^2+1 $ and symmetric case  $N_1 = N_3$, if $N_1 < N_2/2$ the optimal SLEM is equal to $1- ( 2 N_2 ) / ( 4 N_1 N_2 + (N_2 - 1)(N_2 - 2N_1) )$ and the optimal weights are 
$w_2 = w_{-2} = 0$, $w_1 = w_{-1} =    2    / (   4 N_1 N_2 + (N_2 - 1)(N_2 - 2N_1)   )$ and $w_0 = ( N_2 - 2N_1 ) / N_{2}^{2}$
\item For $N = d^2+1 $ and symmetric case  $N_1 = N_3$, if $N_1 \geq N_2/2$ the optimal weights are $w_2 = w_{-2} = 0$, $w_1 = w_{-1} = (2N_{1}-1)/(8N^2_{1}N_{2})$, $w_0 = 0$ and the optimal SLEM 
is $1- 1/(2 N_{1})$.
\end{itemize}
From the optimal weights, it is apparent that for the last symmetric case where $N_1 \geq N_2/2$ the whole topology reduces to a 3-partite graph.
For the nonsymmetric case where $N_1 \neq N_3$, the optimal results are too long to report here.
But interestingly in the nonsymmetric case if $N_1 > N_3$ then the optimal value of the weight $w_{2}$ is zero.

\subsection{FDTQC problem over Complete Graph Topology}
In this subsection we present the complete solution of the FDTQC problem over a quantum network with complete graph topology for all values of $N$.
Since the underlying complete graph is edge transitive
and consequently all of the resultant induced Schreier graphs are edge-transitive \cite{Konstantinova2013}
then the value of optimal weights over all edges (denoted by $w$) are equal.

In the following, we provide some preliminaries on association schemes and group association schemes \cite{js,Ass.sch.,Bose} which have been used for obtaining the spectrum of the induced graphs corresponding to a network with complete underlying graph.

\subsubsection{ Association Schemes }

\begin{definition} {Association Scheme}
\label{AssociationScheme}
\\
Let $V$ be a set of $N$ vertices and assume $\{ R_i : i = 0, 1, \ldots, d \}$ be a set of nonempty relations on $V$ (subsets of $V \times V$ ) satisfying the following conditions, $(1)$ to $(4)$.
Then the pair $Y = \{V, R_i : 0 \leq i \leq d\}$ is called an association scheme.
The constraints on relations are as below, \\
$(1)\;\ \{R_i\}_{0\leq i\leq d}$ is a partition of $V\times V$      \\
$(2)\;\ R_0=\{(\alpha, \alpha) : \alpha\in V \}$              \\
$(3)\;\ R_i=R_i^t$ for $0\leq i\leq d$, where $R_i^t=\{(\beta,\alpha) :(\alpha, \beta)\in R_i\} $       \\
$(4)$ For $(\alpha, \beta)\in R_k$, the number  $p^k_{i,j}=\mid  \{\gamma\in V : (\alpha, \gamma)\in R_i \;\ and \;\ (\gamma,\beta)\in R_j\}\mid$ does not depend on $(\alpha, \beta)$ but only on $i,j$ and $k$.      \\
Then, $Y=(V,\{R_i\}_{0\leq i\leq d})$ defines a symmetric association scheme of class $d$ on $V$.
Further, if $p^k_{ij}=p^k_{ji}$ for all $i,j,k=0,1,\ldots,d$, then $Y$ is called commutative.
\end{definition}

Let $Y=(V,\{R_i\}_{0\leq i\leq d})$ be a commutative symmetric association scheme of class $d$, then the matrices $A_0,A_1,...,A_d$
defined by
\begin{equation}
    \label{adj.}
    \bigl(A_{i})_{\alpha, \beta}\;=\left\{\begin{array}{c}
     1 \quad ,\;\  \mathrm{if} \;(\alpha, \beta)\in R_i, \\
     0 \quad , \;\ \mathrm{otherwise} \quad \quad  \\
    \end{array}\right\},
    \;\ \alpha, \beta \in V
\end{equation}
are adjacency matrices of $Y$ and are such that
\begin{equation}
    \label{ss}
    A_i A_j=\sum_{k=0}^{d}{p_{ij}^kA_{k}}.
\end{equation}
From (\ref{ss}), it is seen that the adjacency matrices $A_0, A_1, \ldots, A_d$ form a basis for a commutative algebra \textsf{A} known
as the Bose-Mesner algebra of $Y$.
This algebra has a second basis $E_0, \ldots, E_d$ primitive idempotents, defined as $E_0 = \boldsymbol{J}/N$, $E_iE_j=\delta_{ij}E_i$, $\sum_{i=0}^d E_i=I$,
where, $N:=|V|$ and $J$ is the $N\times N$ all-$1$ matrix in
$\textsf{A}$.
Let $P$ and $Q$ be the matrices relating the two basis for $\textsf{A}$:     $A_j=\sum_{i=0}^d P_{ij}E_i$ and $E_j=\frac{1}{N}\sum_{i=0}^d Q_{ij}A_i$, for $0\leq j\leq d$.
Then clearly we  have $PQ=QP=NI$.
It also follows that $A_jE_i=P_{ij}E_i$.
The scalars  $P_{ij}$ are called the eigenvalues of the scheme.
Since they are eigenvalues of the matrices $A_j$, they are
algebraic integers (respectively the scalars $Q_{ij}$ are called
the dual eigenvalues of the scheme); the columns of $E_i$ are the
corresponding eigenvectors. Thus $m_i=$ rank($E_i$) is the
multiplicity of the eigenvalue $P_{ij}$ of $A_j$ (provided that
$P_{ij}\neq P_{kj}$ for $k \neq i$). We see that $m_0=1, \sum_i
m_i=N$, and $m_i=$trace$E_i=N(E_i)_{jj}$ (indeed, $E_i$ has only
eigenvalues $0$ and $1$, so rank($E_k$) equals the sum of the
eigenvalues).

One of the important schemes is the group scheme as explained below,

\begin{definition} {Group Association Scheme}
\label{GroupAssociationScheme}
\\
Let $G$ be a group; then it can be proved that the set of relations defined by $R_i  =  \{ (\alpha,\beta): \, \alpha \beta^{-1} \in C_{i} \}$,
where $\{C_i: \, 0 \leq i \leq d\}$ are the set of conjugacy classes of $G$, satisfy the four constraints (1)-(4) in the definition of Association Scheme conditions (Definition \ref{AssociationScheme}), so $Y = \{G;R_i: \, 0 \leq i \leq d\}$ is a symmetric association scheme.
\end{definition}

Defining class sums $\overline{C}_i$ for $i = 0, 1, \ldots , d$ as  $\overline{C}_i  =  \sum_{g \in C_i} {g }  \in \mathbb{C}G$,
%
by the action of $\overline{C_i}$
on group elements in the regular representation we observe that $\forall \alpha, \beta, (\overline{C_i})_{\alpha\beta}  =  (\boldsymbol{A}_i)_{\alpha\beta}$, so $\boldsymbol{A}_i  =  \overline{C_i}  =  \sum_{g \in C_i}{g}$
Thus due to (\ref{ss}), the relation $\overline{C_i} \, \overline{C_j}  =  \sum_{k=0}^{d} { p_{i,j}^{k} \overline{C_k} }$ hold true.
%
%
One can show that $\overline{C_i} \, \overline{C_j}  =  \overline{C_j} \, \overline{C_i}$ therefore the group scheme is a commutative scheme.
Furthermore  they also commute with every element of the group i.e.
\begin{equation}
    \label{eq:2021}
    \begin{gathered}
        \overline{C_i} \, g  =   g  \,  \overline{C_i},   \qquad  \forall i = 0, 1, \ldots , d, \quad \text{and} \quad \forall g \in G
    \end{gathered}
\end{equation}

However from group theory we know that the coefficients $p_{i,j}^{k}$ have to be nonnegative integers, and their value is given by the following relation \cite{Gordon2001}:
\begin{equation}
    \label{eq:1981}
    \begin{gathered}
        p_{i,j}^{k}  =  \frac{|C_i||C_j|}{|G|} \sum_{\chi} {  \frac { \chi(C_i)\chi(C_j) \overline{\chi(C_k)} }  { \chi(1) } }
    \end{gathered}
\end{equation}
where the sum is over all the irreducible characters $\chi$ of $G$ \cite{Tomi32}.
Therefore, the idempotents ${E_0, E_1, \ldots , E_d }$ of the group association scheme $X(G)$ are the projection operators of $\mathbb{C}G$-module, i.e., $E_k  =  \frac  { \chi_k (1) }  { |G| } \sum_{C_i \in G} {\chi_k ( C_{i}^{-1} )C_{i} }$.
Thus, eigenvalues of adjacency matrices of $A_k$ and idempotents $E_k$, respectively, are $P_{ik}  =  \frac{d_i k_k}{m_i} \chi_i (C_k)$ and $Q_{ik}  =  d_k \overline{\chi_k (C_i)}$,
%
where $ d_j = \chi_j(1) $s are the dimensions of the irreducible characters $\chi_j$ which are positive integers,
and $\kappa_i = |C_i|$ is the number of elements in the conjugate class $C_i$, which is the degree of its corresponding adjacency matrix.
Since $\chi_i (g)$ for all $g \in G$ are identical therefore in the formulations above $\chi_i (C_k)$ is employed instead of $\chi_i (g)$.

From relation (\ref{eq:2021}) it is obvious that in any matrix representation the adjacency matrices $A_i$ commute with all elements of group, therefore according to Schur's Lemma $A_i$  should be proportional to identity matrix i.e.
\begin{equation}
    \label{eq:2077}
    \begin{gathered}
        \boldsymbol{A}_i  =  P_{ji} \boldsymbol{I}_{\chi_j(1)},
    \end{gathered}
\end{equation}
where $P_{ji}$ is the $j$-th eigenvalue of the adjacency matrices of $A_i$.

\subsubsection{Association Scheme of Group $S_N$}
\label{sec:AssociationSchemeofGroupSN}

For each partition $n\vdash N$, $S_N$ has a corresponding conjugacy class which consists of those permutations having cycle structure described by $n$.
The cycle structure are the listing of number of cycles of each length (i.e, $n_i$ is the number of $i$ cycles).
We denote by $C_n$ the conjugacy class of $S_N$ consisting of all permutations having cycle structure $n$.
Therefore the number of conjugacy classes of $S_N$, namely the diameter of its scheme is equal to the number of partitions of $N$.

Here we are concerned with the adjacency matrix corresponding to $C_{(2,1,1,1,1...,1)}$ and its spectrum,
since this adjacency matrix is same as the adjacency matrix of the induced graph corresponding to partition $n=(1,1,\ldots,1)$ (Cayley graph) if
the underlying graph of the quantum network is a complete graph.
The adjacency matrix corresponding to $C_{(2,1,1,1,1...,1)}$ is $A_{C_{(2,1,1,\ldots,1)}}  =  A_{C_1} = \sum_{i<j} {\Pi_{ij}}$, 
where $\Pi_{ij}$ is the matrix representation of the transposition $(i \leftrightarrow j)$ in $S_N$-regular representation.

To obtain the spectrum of the adjacency matrix $A_{C_{(2,1,1,\ldots,1)}}$, introduced above, the character of $C_1$ is required which is as below \cite{ri},
\begin{equation}
    \label{eigen2}
    \chi_n (C_1)=\frac{d_n}{\kappa_1}\sum_{j}\left(\left(
    \begin{array}{cc}
        n_j \\ 2
    \end{array}
    \right)
    -
    \left(
    \begin{array}{cc}
        n^{'}_j \\ 2
    \end{array}
    \right) \right).
\end{equation}
Here, $n'$ is the conjugate of partition $n$,
while $n^{'}_j$ and $n_j$ are the $j$-th components of the partitions $n^{'}$ and $n$, and $S_n$ is the irreducible Specht module corresponding to partition $n$.

Since the representation is regular then $m_n $ is equal to $(d_{n})^2$.
$ \kappa_1 = N! / ( (N-1)! 2! ) $ is the cardinality of the first conjugate class $(C_1)$ which is the number of possible transpositions.
Then the eigenvalues of the adjacency matrix $A_{C_{(2,1,1,\ldots,1)}}$ can be written as
\begin{equation}
\label{eigen2220}
P_{n1}  =  \frac{d_{n}\kappa_1}{m_n}\chi_n(C_1)=\sum_{j}\left(\left(
\begin{array}{cc}
 n_j \\ 2
 \end{array}
\right)  -  \left(
\begin{array}{cc}
 n^{'}_j \\ 2
 \end{array}
\right) \right).
\end{equation}

Now using (\ref{eq:2077}), in the irreducible representation corresponding to a given partition $n$, the adjacency matrix $A_{C_{(2,1,1,\ldots,1)}}$  can be written as below
\begin{equation}
\label{eq:2234}
A_{C_{(2,1,1,\ldots,1)}}  =  P_{n1}\cdot \boldsymbol{I}_{\chi_n(1)}  =
\sum_{j}\left(\left(
\begin{array}{cc}
 n_j \\ 2
 \end{array}
\right)  -  \left(
\begin{array}{cc}
 n^{'}_j \\ 2
 \end{array}
\right) \right) \cdot \boldsymbol{I}_{\chi_n(1)} .
\end{equation}

\subsubsection{ Optimal Weights \& SLEM for complete graph topology }

Now we are ready to provide the optimal weights and the SLEM of the discrete time quantum consensus problem over a quantum network with complete graph topology.
To do so, we first obtain the spectrum of the Laplacian matrix $L_n$ of the induced graphs given in (\ref{eq:RobustSolitonDistribution1540}),
which is possible by obtaining the spectrum of $\boldsymbol{A}_{G(n)}$.
This is achievable for partition $n=(1,1,\ldots,1)$, since for this partition we have
\begin{equation}
    \label{eq:2290}
    \begin{gathered}
        \boldsymbol{A}_{G(1,1,\ldots,1)}   =   w   \cdot   \boldsymbol{A}_{C_{(2,1,1,\ldots,1)}},
    \end{gathered}
\end{equation}
where $w$ is the weight on each edge of the complete graph and the spectrum of $\boldsymbol{A}_{C_{(2,1,1,\ldots,1)}}$ is provided above in subsection \ref{sec:AssociationSchemeofGroupSN}.
Restricting relation (\ref{eq:2290}) to an irreducible representation corresponding to partition $n$, using (\ref{eq:LaplacianPartitionBlockSum}) the left hand side of (\ref{eq:2290}) becomes $\tilde{\boldsymbol{A}}_{G(n)}$ while using (\ref{eq:2234}) the right hand side can be written as $w   \cdot  P_{n1}\cdot \boldsymbol{I}_{\chi_n(1)} $.
Therefore, for each block of $L_n$ in (\ref{eq:LaplacianPartitionBlockSum}) we have
\begin{equation}
    \label{eq:2302}
    \begin{gathered}
        \boldsymbol{B}_{n}   =   (W   - P_{n1} ) \cdot   \boldsymbol{I}_{\chi_{n}(1)},
    \end{gathered}
\end{equation}

Considering (\ref{eq:2302}) for the given partition $n$, the eigenvalues of $\boldsymbol{L}_n$ are
\begin{equation}
    \label{2313}
    \begin{gathered}
        \lambda_{n^{'}}(\boldsymbol{L}_n) = W-P_{n^{'}1},
    \end{gathered}
\end{equation}
where $n^{'}$ includes all partitions dominant to partition $n$, with $P_{n^{'}1}$ given in (\ref{eigen2220}).
This is the main result of this section towards calculating the optimal SLEM of the discrete time quantum consensus problem for a quantum network with complete graph topology.

In the following, we provide the optimal results for a quantum network with $N$ qubits.
For complete graph as the underlying graph, the second smallest eigenvalue of the Laplacian matrix $L_n$ for all partitions (other than $n=(N)$) is the eigenvalue of $B_{(N-1,1)}$ in (\ref{eq:LaplacianPartitionBlockSum}) which is $\lambda_2  =  N \cdot w$.
Regarding $\lambda_{max}$ depending on the value of $N$ there are four different categories, namely $N=4l+i$ for $i=0,1,2,3$ and $l \geq 1$.

\begin{itemize}
\item $N=4l$: In this case, since partition  $n=(l,l,l,l)$ is the least dominant partition, $\lambda_{max}$ is obtained from $B_{(l,l,l,l)}$.
From (\ref{eigen2220}), $P_{n1}$ for $n=(l,l,l,l)$ and its conjugate partition $n^{'} = (\underbrace{\scriptstyle 4,4,\ldots,4}_{l})$ can be calculated as $P_{n1}  =  \frac{N(N-16)}{8}$, 
and by substituting $W = wN(N-1)/2$ and $P_{n1}$ in (\ref{2313}) for $n=n^{'}=(l,l,l,l)$ we obtain $\lambda_{max}  =  \frac{3N(N+4)}{8}w$. 
For optimal the weights we have $1-\lambda_2  =  \lambda_{max} - 1$ which results in the optimal $w$ as $w = 16/(N(3N+20))$ 
and thus the corresponding optimal SLEM is $1 - 16/ (3N+20)$.

\item $N=4l+1$: In this case, the optimal weight is $w = 16/(3N^2+20N-15)$ and the optimal SLEM is $1- (16N) / (3N^2+20N-15)$. 

\item $N=4l+2$: In this case, the optimal weight is $w = 16/(3N^2+20N-20)$ and the optimal SLEM is $1- (16N)/(3N^2+20N-20)$. 

\item $N=4l+3$: In this case, the optimal weight is $w = 16/(3N^2+20N-15)$ and the optimal SLEM is $1- (16N)/(3N^2+20N-15)$ 
Interestingly, the optimal results for this category are identical to the second category.

\end{itemize}

For cases other than qubits, i.e. qudits for $d > 2$, the solution procedure is similar as above but for higher number of number of categories that is $(d^2)$.

\section{Conclusion}
\label{sec:Conclusion}

In this paper, we have optimized the convergence rate of the quantum consensus algorithm over a quantum network with $N$ qudits. For the study presented in this paper we have considered the discrete time model of the quantum consensus algorithm.
It is shown that unlike the results obtained for the continuous time model of the algorithm, the convergence rate of the algorithm depends on the value of $d$ is qudits.
By exploiting the Specht module representation of partitions of $N$, we have shown that the spectrum of the Laplacian corresponding to the less dominant partition in the Hasse Diagram includes that of the one level dominant partition.
Using this result, the Aldous' conjecture is generalized to all partitions of integer $N$  and it is shown that the original optimization problem reduces to optimizing the Second Largest Eigenvalue Modulus (SLEM) of the weight matrix.
SLEM depends on two eigenvalues, namely, the second smallest $(\lambda_2(\boldsymbol{L}))$ and the greatest $(\lambda_{max}(\boldsymbol{L}))$ eigenvalues of the Laplacian matrix.
We have proved that $\lambda_2(\boldsymbol{L})$ is same for the induced graphs of all partitions of $N$, while $\lambda_{max}(\boldsymbol{L})$ is obtained from the induced graph of the least dominant and feasible partition of $N$, where the feasiblity depends on the values of $d$ and $N$.
For $N \leq d2$, the $\lambda_{max}(\boldsymbol{L})$ is obtained from partition $(1, 1, \cdots, 1)$ while for larger values of $N$, partition $(1, 1, \cdots, 1)$ is not feasible and $\lambda_{max}(\boldsymbol{L})$ is included in partitions dominant to $(1, 1, \cdots, 1)$.
The semidefinite programming formulation of the FDTQC problem is addressed analytically for $N \leq d^2 + 1$ and a wide range of topologies where closed-form expressions for the optimal convergence rate and the optimal weights are provided.
For the special case of complete graph topology, we have included the complete solution of the FDTQC problem for all values of $N$, where group association schemes are employed for obtaining the spectrum of the induced graphs.


\begin{thebibliography}{99}

\bibitem[\protect\citeauthoryear{Abreu}{Bra}{2007}]%
        {BrazilReview2007}
        {Nair Maria Maia de Abreu}. 
\newblock Old and new results on algebraic connectivity of graphs.
\newblock {\em Linear Algebra and its Applications\/}  {423} (2007), 53--73.








\bibitem[\protect\citeauthoryear{Aldous and Fill}{Aldous and Fill}{2002}]%
        {AldousBook}
{David Aldous} {and} {James~Allen Fill}. 
\newblock Reversible Markov Chains and Random Walks on Graphs.
\newblock
\newblock Unfinished monograph, recompiled 2014, available at http://www.stat.berkeley.edu/$\sim$aldous/RWG/book.html.






\bibitem[\protect\citeauthoryear{Boyd and Vandenberghe}{Boyd and
  Vandenberghe}{2004}]%
        {BoydBook2004}
{Stephen Boyd} {and} {Lieven Vandenberghe}. 
\newblock {\em Convex Optimization}.
\newblock Cambridge University Press, New York, NY, USA.
\newblock





\bibitem[\protect\citeauthoryear{Broadbent and Tapp}{Broadbent and
  Tapp}{2008}]%
        {Broadbent2008}
{Anne Broadbent} {and} {Alain Tapp}. 
\newblock Can Quantum Mechanics Help Distributed Computing?
\newblock {\em SIGACT News\/} {39}, 3 (Sept. 2008), 67--76.
\newblock




\bibitem[\protect\citeauthoryear{Buhrman and R¨ohrig}{Buhrman and
  R¨ohrig}{2003}]%
        {Buhrman2003}
{H. Buhrman} {and} {H. R¨ohrig}. 
\newblock Distributed quantum computing. In {\em
  International Symposium on Mathematical Foundations of Computer Science
  (MFCS), LNCS 2747}. 1--20.
\newblock




\bibitem[\protect\citeauthoryear{Caputo, Liggett, and Richthammer}{Caputo
  et~al\mbox{.}}{2010}]%
        {ProofAldous}
{Pietro Caputo}, {Thomas~M. Liggett}, {and} {Thomas Richthammer}. 
\newblock Proof of Aldous’ spectral gap conjecture.
\newblock {\em Journal American Math Society\/}  {23} (2010), 831--851.
\newblock



\bibitem[\protect\citeauthoryear{Chung}{Chung}{1997}]%
        {Chung1997}
{F.R.K. Chung}. 1997.
\newblock Spectral graph theory.
\newblock {\em American Mathematical Society\/} 
\newblock









\bibitem[\protect\citeauthoryear{Denchev and Pandurangan}{Denchev and
  Pandurangan}{2008}]%
        {Denchev:2008}
{Vasil~S. Denchev} {and} {Gopal Pandurangan}. 
\newblock Distributed Quantum Computing: A New Frontier in
  Distributed Systems or Science Fiction?.
\newblock {\em SIGACT News\/} {39}, 3 (Sept. 2008), 77--95.
\newblock



\bibitem[\protect\citeauthoryear{Fiedler}{Fiedler}{1973}]%
        {Fiedler1973}
{Miroslav Fiedler}. 
\newblock Algebraic connectivity of graphs.
\newblock {\em Czechoslovak Mathematical Journal\/}  {23} (1973).
\newblock









\bibitem[\protect\citeauthoryear{Xiao and Boyd}{Xiao and Boyd}{2004}]%
        {Xiao04Boyd}
{Xiao L.} {and} {Boyd S.}. 
\newblock Fast linear iterations for distributed averaging.
\newblock {\em Systems and Control Letters, vol. 53, pp. 65–78\/} (2004).
\newblock







\bibitem[\protect\citeauthoryear{Horn and Johnson}{Horn and Johnson}{2006}]%
        {Horn2006}
{R.A. Horn} {and} {C.R. Johnson}. 
\newblock Matrix analysis.
\newblock {\em Cambridge University Press\/} (2006).
\newblock


\bibitem[\protect\citeauthoryear{Jadbabaie, Lin, and Morse}{Jadbabaie
  et~al\mbox{.}}{2003}]%
        {Jadbabaie2003}
{A. Jadbabaie}, {Jie Lin}, {and} {A.S. Morse}. 
\newblock Coordination of groups of mobile autonomous agents
  using nearest neighbor rules.
\newblock {\it IEEE Trans. Automat. Control} {48}, 6 (June 2003), 988--1001.
\newblock



\bibitem[\protect\citeauthoryear{Jafarizadeh}{Jafarizadeh}{2015}]%
        {SaberThesis2015}
{Saber Jafarizadeh}. 
\newblock {\em Distributed coding and algorithm optimization for large-scale
  networked systems}.
\newblock Ph.D. Dissertation. University of Sydney, NSW, 2006.
\newblock





\bibitem[\protect\citeauthoryear{Kocarev}{Kocarev}{2013}]%
        {KocarevBookComplexConsensus2013}
{Ljupco Kocarev}. 
\newblock {\em Consensus and Synchronization in Complex Networks}.
\newblock Springer-Verlag, Berlin Heidelberg.
\newblock



\bibitem[\protect\citeauthoryear{Marshall, Olkin, and Arnold}{Marshall
  et~al\mbox{.}}{2011}]%
        {MajorizationBookRefMarshall}
{Albert~W. Marshall}, {Ingram Olkin}, {and} {Barry~C. Arnold}. 
\newblock {\em Inequalities : theory of majorization and its applications}.
\newblock Springer, New York.
\newblock



\bibitem[\protect\citeauthoryear{Mazzarella, Sarlette, and Ticozzi}{Mazzarella
  et~al\mbox{.}}{2013}]%
        {MazzarellaCDC2013}
{L. Mazzarella}, {A. Sarlette}, {and} {F. Ticozzi}. 
\newblock A new perspective on gossip iterations: From
  Symmetrization to quantum consensus. In {\em IEEE 52nd Annual Conference on
  Decision and Control (CDC)}. 250--255.
\newblock




\bibitem[\protect\citeauthoryear{Mazzarella, Sarlette, and Ticozzi}{Mazzarella
  et~al\mbox{.}}{2015}]%
        {PetersenRef15}
{L. Mazzarella}, {A. Sarlette}, {and} {F. Ticozzi}. 
\newblock Consensus for Quantum Networks: Symmetry From
  Gossip Interactions.
\newblock {\it IEEE Trans. Automat. Control} {60}, 1 (Jan 2015), 158--172.
\newblock



\bibitem[\protect\citeauthoryear{Mazzarella, Ticozzi, and Sarlette}{Mazzarella
  et~al\mbox{.}}{2013}]%
        {MazzarellaArXiv}
{L. Mazzarella}, {F. Ticozzi}, {and} {A. Sarlette}. 
\newblock From Consensus to Robust Randomized Algorithms: A Symmetrization Approach.
\newblock {\em quant-ph, arXiv 1311.3364\/} (2013).
\newblock


\bibitem[\protect\citeauthoryear{Jafarizadeh}{Jafarizadeh}{2015}]%
        {SaberContQuanArXiv}
{S. Jafarizadeh}. 
\newblock Optimizing the Convergence Rate of the Continuous Time Quantum Consensus.
\newblock {\em cs-SY, arXiv 1509.05823\/} (2015).
\newblock



\bibitem[\protect\citeauthoryear{Olfati-Saber and Murray}{Olfati-Saber and
  Murray}{2004}]%
        {Olfati2004}
{Reza Olfati-Saber} {and} {Richard~M. Murray}. 
\newblock Consensus Problems in Networks of Agents With Switching Topology and Time-Delays.
\newblock {\it IEEE Trans. Automat. Control} {49}, 9 (September 2004),
  1520--1533.
\newblock



\bibitem[\protect\citeauthoryear{Jafarizadeh and Jamalipour}{Jafarizadeh and
  Jamalipour}{2011}]%
        {SaberSensorJournal2011}
{S. Jafarizadeh} {and} {A. Jamalipour}. 
\newblock Fastest distributed consensus problem on fusion of two star sensor networks.
\newblock {\it IEEE Sensors Journal} {11}, 10 (October 2011), 2494--2506,
  1520--1533.
\newblock


\bibitem[\protect\citeauthoryear{Jafarizadeh and Jamalipour}{Jafarizadeh and
  Jamalipour}{2011}]%
        {SaberICC2011}
{S. Jafarizadeh} {and} {A. Jamalipour}. 
\newblock Fastest distributed consensus on star-mesh hybrid sensor networks.
\newblock {\it IEEE International Conference on Communications (ICC)} (June 2011), 1--5,
  1520--1533.
\newblock


\bibitem[\protect\citeauthoryear{Ren and Beard}{Ren and Beard}{2008}]%
        {WeiBookVehicleConsensus2008}
{Wei Ren} {and} {Randal~W. Beard}. 
\newblock {\em Distributed consensus in multi-vehicle cooperative control: theory and applications}.
\newblock Springer, London.
\newblock



\bibitem[\protect\citeauthoryear{S., Linial, and Widgerson}{S.
  et~al\mbox{.}}{2006}]%
        {Hoory2006}
{Hoory S.}, {N. Linial}, {and} {A. Widgerson}. 
\newblock Expander graphs and their applications.
\newblock {\it Bull. Amer. Math. Soc.} {43}, 4 (October 2006), 439--531.
\newblock


\bibitem[\protect\citeauthoryear{Sagan}{Sagan}{2001}]%
        {Segan2001}
{Bruce Sagan}. 
\newblock {\em The Symmetric Group: Representations, Combinatorial Algorithms,
  and Symmetric Functions}.
\newblock Springer, New York.
\newblock



\bibitem[\protect\citeauthoryear{James, Liebeck}{James  et~al\mbox{.}}{2001}]%
        {Gordon2001}
{Gordon James}, {Martin Liebeck}. 
\newblock {\em Representations and Characters of Groups}.
\newblock Cambridge University Press.
\newblock



\bibitem[\protect\citeauthoryear{James}{James}{1978}]%
        {Gordon1978}
{Gordon James}. 
\newblock {\em The Representation Theory of the Symmetric Group}.
\newblock Spreinger-Verlag, Berlin.
\newblock




\bibitem[\protect\citeauthoryear{Shi, Dong, Petersen, and Johansson}{Shi
  et~al\mbox{.}}{2015a}]%
        {Petersen2015IEEETranAutControl}
{G. Shi}, {D. Dong}, {I. Petersen}, {and} {K. Johansson}. 
\newblock Reaching a Quantum Consensus: Master Equations that Generate Symmetrization and Synchronization.
\newblock {\it IEEE Trans. Automat. Control} (2015), 1--14.
\newblock



\bibitem[\protect\citeauthoryear{Shi, Fu, and Petersen}{Shi
  et~al\mbox{.}}{2015b}]%
        {Petersen2015ACCPartI}
{Guodong Shi}, {Shuangshuang Fu}, {and} {Ian~R. Petersen}. 
\newblock Quantum network reduced-state synchronization part
  I-convergence under directed interactions. In {\em American Control
  Conference (ACC), 2015}.
\newblock


\bibitem[\protect\citeauthoryear{Shi, Fu, and Petersen}{Shi
  et~al\mbox{.}}{2015c}]%
        {Petersen2015ACCPartII}
{Guodong Shi}, {Shuangshuang Fu}, {and} {Ian~R. Petersen}. 
\newblock Quantum network reduced-state synchronization part
  II-the missing symmetry and switching interactions. In {\em American Control
  Conference (ACC), 2015}. 92--97.
\newblock



\bibitem[\protect\citeauthoryear{Tsitsiklis, Bertsekas, and Athans}{Tsitsiklis
  et~al\mbox{.}}{1986}]%
        {Tsitsiklis1986}
{J.N. Tsitsiklis}, {D.P. Bertsekas}, {and} {M. Athans}. 
\newblock Distributed asynchronous deterministic and
  stochastic gradient optimization algorithms.
\newblock {\it IEEE Trans. Automat. Control} {31}, 9 (Sep 1986), 803--812.
\newblock




\bibitem[\protect\citeauthoryear{Ingram}{Ingram}{1950}]%
        {ri}
{R. Ingram}. 
\newblock Some characters of the symmetric group. In {\em Proc. Amer. Math. Soc.}. 1, 358-369.
\newblock





\bibitem[\protect\citeauthoryear{Jafarizadeh and Sufiani}{Jafarizadeh
  et~al\mbox{.}}{2008}]%
        {js}
{M. A. Jafarizadeh},  {R. Sufiani},  {and} {S. Jafarizadeh}. 
\newblock Calculating effective resistances on underlying networks of association schemes.
\newblock {\it Journal of Mathematical Physics } 49 (7), (2008).
\newblock




\bibitem[\protect\citeauthoryear{Konstantinova}{Konstantinova}{2013}]%
        {Konstantinova2013}
{Elena Konstantinova}. 
\newblock Some problems on Cayley graphs.
\newblock {\it University of Primorska Press  } (2013).
\newblock





\bibitem[\protect\citeauthoryear{Bailey}{Bailey}{2004}]%
        {Ass.sch.}
{R. A. Bailey}. 
\newblock Association Schemes: Designed Experiments, Algebra and Combinatorics.
\newblock {\it Cambridge University Press, Cambridge  } (2004).
\newblock




\bibitem[\protect\citeauthoryear{Bose and Shimamoto}{Bose
  et~al\mbox{.}}{1952}]%
        {Bose}
{R. C. Bose and T. Shimamoto}. 
\newblock Classification and analysis of partially balanced incomplete block designs with two associate classes.
\newblock {\it J. Amer. Statist. Assoc.   } (1952), 47, 151 .
\newblock



\bibitem[\protect\citeauthoryear{Tomiyama}{Tomiyama}{1999}]%
        {Tomi32}
{Masato Tomiyama}. 
\newblock Characterization of the Group Association Scheme of PSL(2, 7).
\newblock {\it Journal of Combinatorial Theory, Series A   } (1999), Vol. 88, Issue 2, pp 306–341 .
\newblock



\bibitem[\protect\citeauthoryear{Nielsen and Chuang}{Nielsen
  et~al\mbox{.}}{2002}]%
        {Nielsen}
{M. A. Nielsen and I. L. Chuang}. 
\newblock Quantum Computation and Information.
\newblock {\it Cambridge, U.K.: Cambridge Univ. Press   } (2002).
\newblock



\bibitem[\protect\citeauthoryear{Kraus}{Kraus}{1983}]%
        {Kraus}
{K. Kraus}. 
\newblock Effects, and Operations: Fundamental Notions of Quantaum Theory.
\newblock {\it Berlin, Germany: Springer-Verlag, ser. Lecture notes in Physics.   } (1983).
\newblock



\bibitem[\protect\citeauthoryear{Mendl and Wolf}{Mendl
  et~al\mbox{.}}{2009}]%
        {Mendl}
{C. B. Mendl and M. M. Wolf}. 
\newblock Unital quantum channels—Convex structure and revivals of birkhoff’s theorem.
\newblock {\it Commun. Math. Phys., vol. 289, pp. 1057–1096,} (2009).
\newblock




\end{thebibliography}
\end{document}